\documentclass[nofootinbib,aps,twocolumn,superscriptaddress,showpacs,english]{revtex4-1}

\usepackage[T1]{fontenc}
\usepackage[utf8]{inputenc}
\usepackage[english]{babel}
\usepackage{amsmath}
\usepackage{amssymb}
\usepackage{wasysym}
\usepackage{graphicx}
\usepackage{xcolor}
\usepackage{siunitx}
\usepackage{bm} 
\usepackage{cancel}
\usepackage{mdframed}
\usepackage{wrapfig}
\usepackage{braket}
\usepackage{todonotes}
\usepackage{pseudocode}

\usepackage{xcolor} 
\usepackage{pst-node} 

\renewcommand{\vec}[1]{{\bf #1}}
\newcommand{\bfA}{{\bf A}}
\newcommand{\bfC}{{\bf C}}
\newcommand{\bfG}{{\bf G}}
\newcommand{\opn}{{\mathop{\hat{n}}}}
\newcommand{\bfalpha}{\bm \alpha}
\newcommand{\bfsigma}{\Sigma}
\newcommand{\opH}{{\mathop{\hat{H}}}}
\newcommand{\opa}{{\mathop{\hat{a}}}}
\newcommand{\opc}{{\mathop{\hat{c}}}}

\newcommand{\myhat}[1]{#1^{*}}
\newcommand{\CS}{\,\boldsymbol \vert\,}
\newcommand{\im}{{\mathrm{i}}}
\newcommand{\BK}{{\cal I}}
\newcommand{\limOm}{\Omega}

\newcommand{\GrazTh}{Institute of Theoretical and Computational Physics, Graz University of Technology, NAWI Graz, 8010 Graz, Austria}

\begin{document}

\title[Bayesian parametric analytic continuation of Green's functions]{Bayesian parametric analytic continuation of Green's functions}

\author{Michael Rumetshofer}  
\email{m.rumetshofer@tugraz.at} 
\affiliation{\GrazTh} 
\author{Daniel Bauernfeind}    
\affiliation{\GrazTh}
\author{Wolfgang von der Linden}    
\affiliation{\GrazTh}

\date{\today}

\begin{abstract}
Bayesian parametric analytic continuation (BPAC) is proposed for the analytic continuation of noisy imaginary-time Green's function data as, e.g., obtained by continuous-time quantum Monte Carlo simulations (CTQMC). Within BPAC, the spectral function is inferred from a suitable set of parametrized basis functions.
Bayesian model comparison then allows to assess the reliability of different parametrizations.
The required evidence integrals of such a model comparison are determined by nested sampling. 
Compared to the maximum entropy method (MEM), routinely used for the analytic continuation of CTQMC data, the presented approach allows to infer whether the data support specific structures of the spectral function.
We demonstrate the capability of BPAC in terms of CTQMC data for an Anderson impurity model (AIM) that shows a generalized Kondo scenario and compare the BPAC reconstruction to the MEM, as well as to the spectral function obtained from the real-time fork tensor product state impurity solver, where no analytic continuation is required.
Furthermore, we present a combination of MEM and BPAC and its application to an AIM arising from the {\it ab initio} treatment of SrVO$_3$.
\end{abstract}

\maketitle

\section{Introduction}

In quantum many-body physics, Green's functions are often calculated in the imaginary-time domain, e.g., by numerical approaches, such as quantum Monte Carlo \cite{Hirsch_QMC_1986,Foulkes_QMC_2001,Werner_CTQMC_2006,Gull_CTQMC_2011}.
The imaginary-time Green's function $G(\tau)$ is related to the spectral function $A(\omega)$ by a Laplace transform, 
the so-called analytic continuation (AC).
Obtaining $A(\omega)$ from  $G(\tau)$ corresponds to inverting a Fredholm integral of the first kind and small changes in $G(\tau)$ correspond to large differences in $A(\omega)$.
Therefore, the inversion of this problem is highly ill posed and very unstable against numerical noise, and even errors at the level of machine precision can lead to unphysical results in practice.

Many different methods to perform the AC have been proposed, e.g., series expansions such as the Padé method \cite{FERRISPRABHU_Pade_1973},
machine learning \cite{Arsenault_machinelearning_2017},
stochastic methods \cite{White_AverSpectrumMethod_1991,Sandvik_StochAC_1998,Mishchenko_samplingMaxEnt_2000,Beach_sochasticAC_2004,Fuchs_samplingMaxEnt_2010,Sandvik_StochAC_2016,Ghanem_StochS_2017},
and the maximum entropy method (MEM) \cite{Silver_CTQMCMaxEnt_1990,Gubernatis_QMCMaxEnt_1991,Skilling_MaxEnt_1984,Bao_optimMaxEnt_2016,Kraberger_MaxEnt_2017}.
The latter is a consistent approach as it is based on Bayesian probability theory; however, a highly ignorant entropic prior is used, which merely accounts for positivity and additivity of the reconstructed signal. 

Here, we propose a physically motivated prior that takes into account the knowledge
of typical structures of a spectral density, which results in a parametric instead of a form-free reconstruction. 
The Bayesian parametric analytic continuation (BPAC) is based 
on Bayesian parameter estimation \cite{vonderLinden_Bayes_2014,Gregory_Bayes_2005,Jaynes_LogicOfSience_2003,Sivia_DataAnalysis_1996} to obtain a parametrized spectral function.
To be precise, we use asymmetric Lorentzians as well as suitable tails to build up the spectral function.
To validate parametrizations, we use Bayesian model comparison. The required evidence integrals 
are computed by nested sampling (NESA) \cite{Skilling_NESA_2004}. 
With this approach, we can compare parametrizations, e.g., with a different number of asymmetric Lorentzians.
Compared to the other methods, this allows to ask specific questions about the spectral function, e.g., about the reliability of peaks in the spectral function.

We demonstrate the capability of BPAC in terms of the spectral function of an Anderson impurity model (AIM), which exhibits a generalized Kondo scenario.
We calculate the imaginary-time Green's function of the AIM with continuous-time quantum Monte Carlo (CTQMC) \cite{Werner_CTQMC_2006,triqs_2015,Seth_CTHYB_2016} and compare the spectral functions, obtained from BPAC with that of a MEM reconstruction.
The MEM spectrum shows a peak close to the Abrikosov-Suhl resonance, but it is unclear whether this feature is physical or an artifact of the AC.
BPAC can answer this question, showing that it is, in fact, an artifact. Additionally, we successfully compare the BPAC result to the solution obtained with the recently developed real-time fork tensor product state (FTPS) impurity solver \cite{Bauernfeind_FTPS_2017,Bauernfeind_THESIS_2018}, which directly computes the spectral density, without any AC.

In addition, we present a combination of MEM and BPAC and its application to an AIM arising from the dynamical mean-field theory (DMFT)~\cite{MetznerVollhardt_Dinf,Georges_DMFToriginal} treatment of SrVO$_3$. The spectral function obtained with the real-time FTPS solver shows a three-peak structure in the upper Hubbard band, which is absent in the CTQMC + MEM spectral function. We investigate the question whether the absence of this structure is due to a failure of MEM or due to the ill-posed nature of the AC.

{
The applications of BPAC presented in this paper demonstrate that BPAC is a valuable addition to nonparametric reconstruction methods, such as MEM,
e.g., to assess whether the data support specific features found in the MEM reconstruction.
}

The paper is organized as follows:
We first introduce the AC problem in Sec. \ref{sec:AC}.
In Sec. \ref{sec:Parametrizations}, we define and evaluate our parametrizations of the spectral function.
Bayesian parameter estimation and model comparison are discussed in Sec. \ref{sec:ParameterEstimation}.
Finally, in Secs. \ref{sec:Application} and \ref{sec:application2}, we demonstrate the capability of BPAC, first, on an AIM that exhibits a generalized Kondo scenario and, second, on an AIM stemming from the {\it ab initio} treatment of SrVO$_3$.


\section{Analytic continuation and maximum entropy methods}
\label{sec:AC}

Dynamical correlation functions in imaginary time as obtained from CTQMC, 
obey the (anti-) periodicity relation $G(\tau+\beta) = \mp G(\tau)$. The upper sign (-) holds for fermions, and the lower sign (+) holds for bosons.
Due to (anti-) periodicity, $G(\tau)$ is uniquely determined by its values in the interval $\tau\in[0,\beta)$, and its discrete Fourier representation is
\begin{align}
 G(\tau) = \frac{1}{\beta}\sum_{\omega_n} e^{-\im\omega_n\tau}\mathcal{G}(\im\omega_n) \;.
 \label{eq:FT}
\end{align}
The sum is over the Matsubara frequencies $\omega_n=(2n+1)\pi/\beta$ for fermions and $\omega_n=2n\pi/\beta$ for bosons, where $n\in\mathbb{Z}$.
The retarded Green's function $\mathcal{G}(\omega+\im 0^+)$ and Matsubara Green's function $\mathcal{G}(\im\omega_n)$ are related through the analyticity of $\mathcal{G}(z)$.
The spectral function $A(\omega) = -\frac{1}{\pi} \Im \mathcal{G}(\omega+i0^+)$ determines
\begin{align}
  \mathcal{G}(z) = \mp \int d\omega\; \frac{A(\omega)}{z-\omega} \;.
  \label{eq:KK}
\end{align}
Merging Eqs. \ref{eq:FT} and \ref{eq:KK} produces the relation between the imaginary-time Green's function $G(\tau)$ and the spectral function $A(\omega)$,
e.g., for fermions,
\begin{align}\label{eq:A:omega}
  G(\tau) =   \int d\omega\; {\frac{e^{-\omega\tau}}{e^{-\beta\omega}+1}} A(\omega) = \int d\omega\; {K(\tau,\omega)} A(\omega) \;.
\end{align}
%
To handle the problem numerically, we discretize the functions $G(\tau)$ and $A(\omega)$, i.e., $(\bfG)_n = G_n=G(\tau_n)$ and $(\bfA)_m = A_m=A(\omega_m)$.
Consequently, discretizing the kernel $K_{nm}=K(\tau_n,\omega_m)$ produces the matrix equation $\bfG = K \bfA$.
Note that, as shown by Ref. \cite{Ghanem_StochS_2017}, choosing the discretization grid already includes prior information and is equivalent to imposing a default model. Here, we restrict ourselves to linear discretization grids.

The determination of $\bfG$ from $\bfA$ is straight forward, but the inversion $\bfA=K^{-1}\bfG$ is a highly {ill-posed} inversion problem, which is impossible to tackle without taking the noise statistics and reliable prior knowledge consistently into account.
In assuming a multivariate normal distribution with the covariance matrix $\Sigma$ of the QMC data vector $\bfG_\text{d}$, the maximum likelihood (ML) estimator $\bfA_\text{ML}$ is obtained by minimization of $\chi^2(\bfA) = (K\bfA-\bfG_\text{d})^T \Sigma^{-1}(K\bfA-\bfG_\text{d})$.
Due to the {ill posedness} of the problem, $\bfA_\text{ML}$ is, in general, not a satisfying solution, e.g., negative, spiky, and unnormalized.
Additional information, e.g., positivity, smoothness, etc., can be incorporated to regularize the problem.
As shown by Skilling \cite{Skilling_MaxEnt_1989} on a rigorous probabilistic footing, introducing an entropy term,
\begin{align}
 S(A) = \int d\omega\; \left( A(\omega) - D(\omega) - A(\omega) \ln\frac{A(\omega)}{D(\omega)} \right)
\end{align}
and maximizing $-\frac{1}{2}\chi^2(\bfA) + \alpha S(\bfA)$, where $S(\bfA)$ is the discretized version of the entropy $S(A)$, regularizes the problem.
In this so-called MEM, the standard model $D(\omega)$ determines the prior
information about the spectral function, and the hyperparameter $\alpha$, roughly speaking, determines the mixing ratio between the ML solution and the standard model $D(\omega)$.
A small $\alpha$ produces the ML solution, whereas for large $\alpha$, the spectral function approaches the standard model $D(\omega)$. 
The hyperparameter $\alpha$ can be adjusted in various ways, e.g., historic MEM \cite{Gull_MaxEnt_1978,Skilling_MaxEnt_1984}, classical MEM \cite{Gull_MaxEnt_1989,fischer_importance_1996}, and Bryan MEM \cite{Bryan_MaxEnt_1990}.


\section{Parametrization of spectral functions}
\label{sec:Parametrizations}

In the present paper, we propose BPAC.
This approach circumvents the ill-posed problem to a large extend by representing the spectrum using only a few parameters.

We want to build up the spectral function as a sum of peaks where each peak is supposed to correspond to a real peak in the spectrum, e.g., the Abrikosov-Suhl resonance, the left and right Hubbard band, etc.
Due to the natural line width of spectral lines, the obvious choice is to use Lorentzian functions.
General peaks in spectral functions are not single Lorentzians and can show shoulders or plateaus, e.g., between a Hubbard band and the Abrikosov-Suhl resonance.
Therefore, we introduce a sum of asymmetric Lorentzian functions for each peak and add additional tails to describe the decay of the spectrum at higher energies.
It depends on the desired accuracy $|A(\omega)-A_0(\omega)|$ of the reconstructed spectrum $A(\omega)$ to the true spectral function $A_0(\omega)$, whether using one or more Lorentzians per peak is more appropriate.
In our parametrization, the $n$th peak is located at position $\mu_{n}$ and is built up by $C_{n}$ asymmetric Lorentzians, each $i\in\{1,2,...,C_n\}$ having its own individual amplitude $a^i_n$, left width $\gamma^{i,\text{l}}_n$, and right width $\gamma^{i,\text{r}}_n$, resulting in  $(3C_n+1)$ parameters
$
 {\tilde\alpha }_n = \{\mu_n,\{a^i_n,\gamma^{i,\text{l}}_n,\gamma^{i,\text{r}}_n\}\}
$,
\begin{align}
   f_n(\omega|\tilde\alpha_n) = \begin{cases}
   \sum\limits_{i=1}^{C_n}a_n^i\frac{2\gamma^{i,\text{l}}_n}{\gamma^{i,\text{l}}_n + \gamma^{i,\text{r}}_n} L(\omega|\mu_n,\gamma^{i,\text{l}}_n) &\text{for~} \omega<\mu_n \\
    \sum\limits_{i=1}^{C_n}a_n^i\frac{2\gamma^{i,\text{r}}_n }{\gamma^{i,\text{l}}_n + \gamma^{i,\text{r}}_n} L(\omega|\mu_n,\gamma^{i,\text{r}}_n)&\text{for~} \mu_n\leq\omega
\end{cases} \;.
\label{eq:parametrization}
\end{align}
$L(\omega|\mu,\gamma)$ denotes the normalized Lorentzian with center $\mu$ and width $\gamma$,
\begin{align}
L(\omega|\mu,\gamma) &:= \frac{1}{\pi}\frac{\gamma}{(\omega-\mu)^{2}+\gamma^{2}}\;.
\end{align}
%
%
We refer to the case of $C_n>1$ as {\it split} Lorentzian.
The prefactors ${2\gamma^{i,\text{l}}_n }{(\gamma^{i,\text{l}}_n + \gamma^{i,\text{r}}_n)^{-1}}$ and ${2\gamma^{i,\text{r}}_n }{(\gamma^{i,\text{l}}_n + \gamma^{i,\text{r}}_n)^{-1}}$ in Eq. \ref{eq:parametrization} ensure continuity and normalization of the asymmetric Lorentzians.
The spectrum in the frequency interval $I_{\limOm}=[\limOm^\text{l},\limOm^\text{r}]$ 
is described by a superposition of the Lorentzians $f_n(\omega|\tilde\alpha_n)$. For the tails
of the spectrum outside the interval $I_{\limOm}$, a power-law decay is more appropriate.
Then, the total spectral function becomes
\begin{align}\label{eq:A_parametrization}
  A(\omega|\boldsymbol{\alpha}) = \begin{cases}
                                    a^\text{l} \; {|\omega-{\mu^\text{l}}|^{-\nu^\text{l}}}  & \text{for~} \omega < \limOm^\text{l} \\
                                    \sum_{n=1}^{N_\text{p}}\; f_n(\omega|\tilde\alpha_n) & \text{for~} \limOm^\text{l} \leq \omega \leq \limOm^\text{r} \\
                                    a^\text{r} \; {|\omega-{\mu^\text{r}}|^{-\nu^\text{r}}}  & \text{for~} \limOm^\text{r} < \omega
                                  \end{cases}\;.
\end{align}
$N_\text{p}$ is the total number of peaks and $\boldsymbol{\alpha}$ includes the parameters of the Lorentzians $\boldsymbol{\tilde\alpha}$ and the six parameters of the tails, namely, $\{\mu^\text{l}, \nu^\text{l} , \limOm^\text{l},\mu^\text{r}, \nu^\text{r} , \limOm^\text{r}\}$.
The parameters $a^\text{l}$ and $a^\text{r}$ are determined by forcing the spectral function to be continuous.
Hence, we end up with a spectral function described by $N_\alpha = \left( \sum_n^{N_\text{p}}(3C_n+1) \right) + 6$ parameters.

In a first test, we analyze how well parametrized spectral functions using Lorentzians of increasing complexity as defined in Eq. \ref{eq:A_parametrization} can represent typical physical spectra.
As a test case, we use the spin-down and spin-up spectra $A(\omega)$ of the AIM discussed in Sec. \ref{sec:Application}.
\begin{figure}[h]
\begin{center}
  \vspace{0.5cm}
  \flushleft(a)\\\includegraphics[width=1.0\columnwidth,angle=0]{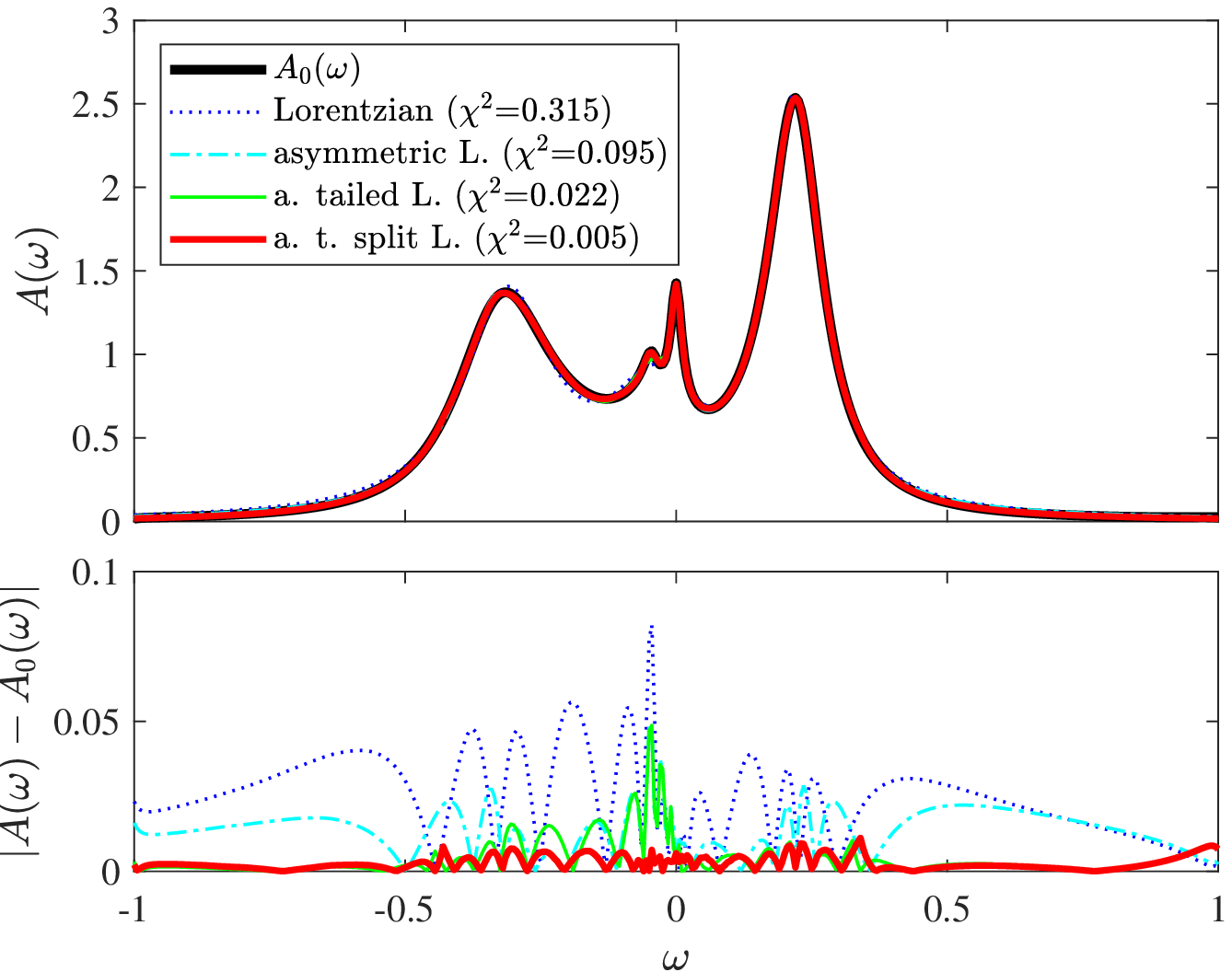}
  (b)\\\includegraphics[width=1.0\columnwidth,angle=0]{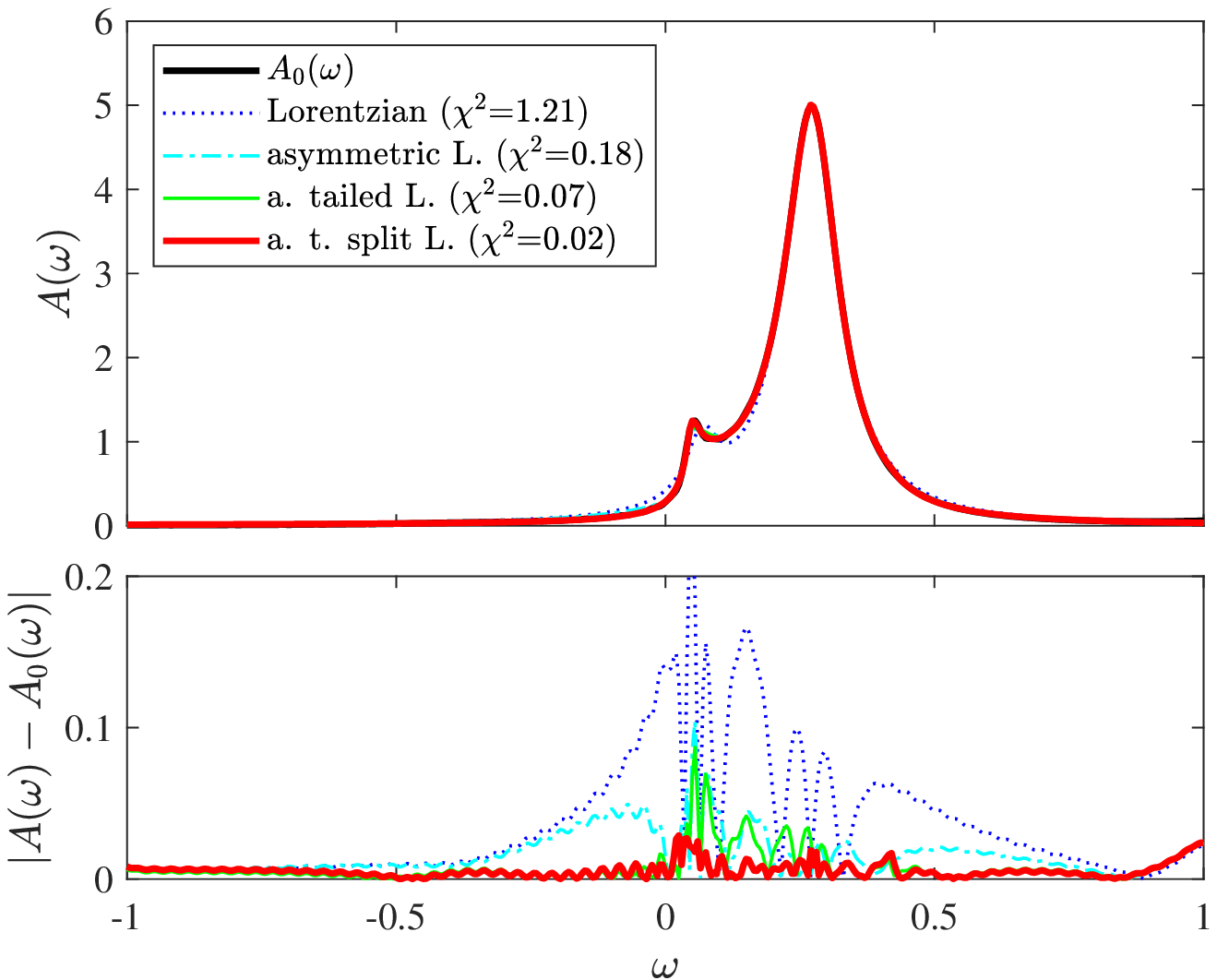}
  \caption{(a) Spin-down and (b) spin-up spectral function $A_0(\omega)$ of the AIM presented in Sec. \ref{sec:Application} and obtained with the FTPS impurity solver.
  The upper plots show fits using Lorentzians (labeled: Lorentzian), asymmetric Lorentzians (labeled: asymmetric L.), asymmetric Lorentzians including tails (labeled: a. tailed L.), and asymmetric and split Lorentzians ($C_n=2$, $\forall n$) including tails (labeled: a. t. split L.).
  We refer to the main text for the detailed definition of the parametrized spectral functions.
  The quadratic errors $\chi^2$ and the deviation $|A(\omega)-A_0(\omega)|$ (lower plots) indicate the increasing quality of the fit.}
  \label{fig:fitA}
\end{center}
\end{figure}
Fig. \ref{fig:fitA} shows the (reference) spectra obtained by the FTPS impurity solver ($A_0(\omega)$, solid black lines) for the (a) spin-down and the (b) spin-up electrons and compares $A_0(\omega)$ to approximations of increasing complexity with the generalized Lorentzian ansatz (colored lines).
For the (a) spin-down spectrum, we used a four-peak ($N_\text{p}=4$) spectral function whereas the (b) spin-up spectrum is approximated by a two-peak ($N_\text{p}=2$) spectral function.
We determine the parameters in Eq. \ref{eq:A_parametrization} by the least-squares approach.
We find that using asymmetric Lorentzians instead of symmetric ones, allows to describe the peaks much better, whereas including tails leads to visible improvements in the high-energy regions, see the lower plots in Figs. \ref{fig:fitA} (a) and \ref{fig:fitA} (b).
Using split Lorentzians further decreases the deviation to the reference spectrum.
Note that the small oscillations in the lower plots of Fig. \ref{fig:fitA} (a) and (b) are artifacts caused by the Fourier transformation of the finite-time solution of the FTPS impurity solver.

To summarize, we find that the parametrization of Eq.~\ref{eq:A_parametrization} is highly flexible and allows to represent reliably the entire structure of the spectrum.


\section{Parameter estimation and model comparison}
\label{sec:ParameterEstimation}

\subsection{Bayesian data analysis}

In this section, we discuss how to determine the parameters $\bfalpha$ of the spectral function $\bfA(\bfalpha)$ and how to judge which parametrization is supported best by the data.
From now on, we call a parametrization of the spectral function the {model} $M=M(\bfC)$, depending implicitly on the peak complexity $\bfC$, which also includes the number of peaks $N_\text{p}=\dim(\bfC)$.
Note that $M$ does not define the values of the model parameters. Bayes' theorem gives
\begin{widetext}
 \begin{align}
  \underbrace{p(\bfalpha |\bfG_\text{d},\bfsigma,M,\BK)}_{\text{posterior}} \; \underbrace{p(\bfG_\text{d}|M,\bfsigma,\BK)}_{\text{data evidence}} &=   {\underbrace{p(\bfG_\text{d}|\bfalpha,\bfsigma,M,\BK)}_{\text{likelihood}}\;\underbrace{p(\bfalpha|M,\BK)}_{\text{prior}} }\;,
\end{align}
\end{widetext}
%
%
where  $\bfG_\text{d}$ stands for the $N_\text{d}$ imaginary-time data points from CTQMC.
We assume that they have a multivariate Gaussian stochastic error. The corresponding
covariance matrix is denoted by  $\Sigma$. The kernel $K$ is included in the conditional complex $\BK$.
The likelihood is therefore,
\begin{align}
  {L}(\boldsymbol{\alpha}) :&= p(\bfG_\text{d}|\boldsymbol{\alpha},\bfsigma,M,\BK) 
  = \frac{1}{\sqrt{({2\pi})^{N_\text{d}} \det (\Sigma)}} e^{-\frac{1}{2}\chi^2(\boldsymbol{\alpha})} \notag \\
  \chi^{2}(\boldsymbol{\alpha}) &=
  \left( K {\bf A}(\bfalpha ) - \bfG_\text{d}\right) ^T \Sigma^{-1} \left( K {\bf A}(\bfalpha ) - \bfG_\text{d}\right)  \;.
  \label{eq:likelihood}
\end{align}
Since correlations are negligible in the data sets used in the present paper we take $\Sigma_{ij}= \sigma_i^2 \delta_{ij}$ in the following.

We use the prior probability to restrict the parameter space for two reasons: First, to only obtain physical results, e.g., by forcing the spectral function to be positive ($a_n^i>0$, $\forall n,i$);
second, to build in and test additional knowledge about the spectral function, e.g., by forcing a peak to appear in a chosen energy interval to analyze whether the data supports this peak. 
For example, a question that can be answered by BPAC could be: {\it Is there a side peak left to the Abrikosov-Suhl resonance?}
Bayesian model comparison allows to judge whether the model including the additional peak is more probable than the model without this peak.
Apart from the above restrictions, we used a flat prior,
\begin{align}
 \pi(\boldsymbol{\alpha}) &:= p(\bfalpha|M,\BK) \notag \\
 &= \prod_{i=1}^{N_\alpha} \; \frac{1}{\alpha^\text{max}_i-\alpha^\text{min}_i}\;\Theta(\alpha^\text{min}_i<\alpha_i<\alpha^\text{max}_i) \;.
\label{eq:prior}
\end{align}
In the present paper, we restrict ourselves to this prior, which is simple to implement and computationally inexpensive since sampling from uniform distributions is cheap. More advanced priors are possible although, e.g., including transformation invariance, smoothness, using testable information, such as the normalization of the spectral function, or even using the entropic prior, may help to improve the results.

Primarily, we are interested in the probability density for the spectral function $p(\bfA |\bfG_\text{d},\bfsigma,M,\BK)$, which we easily obtain from the posterior distribution $p(\bfalpha |\bfG_\text{d},\bfsigma,M,\BK)$ by using the marginalization rule,
\begin{align}
    p(\bfA |\bfG_\text{d},\bfsigma,M,\BK) = \int d\bfalpha \; \underbrace{p(\bfA |\bfalpha,M,\BK)}_{\delta(\bfA - \bfA(\bfalpha))} p(\bfalpha |\bfG_\text{d},\bfsigma,M,\BK) \;.
    \label{eq:p(A)}
\end{align}
Additionally, we want to calculate the data evidence $p(\bfG_\text{d}|\bfsigma,M,\BK)$, which allows to assess the probability of different models relative to each other since the probability for model $M$ is proportional to the data evidence,
\begin{align}
  P(M|\bfG_\text{d},\bfsigma,\BK) =   \frac{1}{p(\bfG_\text{d}|\BK)} \; p(\bfG_\text{d}|\bfsigma,M,\BK)\;P(M|\BK)\;.
\end{align}
In the so-called odds ratio, the ratio between the probabilities for models $M_1$ and $M_2$, the unknown probability $p(\bfG_\text{d}|\BK)$ cancels out, and we get
\begin{align}
 \mathcal{O} = \frac{P(M_1|\bfG_\text{d},\bfsigma,\BK)}{P(M_2|\bfG_\text{d},\bfsigma,\BK)} = \underbrace{\frac{p(\bfG_\text{d}|M_1,\bfsigma,\BK)}{p(\bfG_\text{d}|M_2,\bfsigma,\BK)}}_{\text{Bayes factor}} \underbrace{\frac{P(M_1|\BK)}{P(M_2|\BK)}}_{\text{prior odds}} \;.
\end{align}
In the present paper, we set the prior odds to one since we do not want to favor any model.


\subsection{Evaluating posterior and data evidence}

In this section, we want to give a very brief introduction to NESA, which is a method providing both the data evidence and the samples from the posterior.
We provide the basic equations in this section but refer to Refs. \cite{Sivia_DataAnalysis_1996,Skilling_NESA_2006,vonderLinden_Bayes_2014} for the detailed derivation of NESA.

Skilling \cite{Skilling_NESA_2004} proposed to write the data evidence integral as the Lebesgue integral,
\begin{align}\label{eq:eveidence}
p(\bfG_\text{d}|M,\bfsigma,\BK) &= 
 \int d\boldsymbol{\alpha} \;
     \underbrace{p(\bfG_\text{d}|\boldsymbol{\alpha},\bfsigma,M,\BK)}_{{L}(\boldsymbol{\alpha})} \underbrace{p(\boldsymbol{\alpha}|M,\BK)}_{\pi(\boldsymbol{\alpha})} \notag\\
     &=  \int d\lambda \; X(\lambda)\;,
\end{align}
where the integral over the prior mass,
\begin{align}
 X(\lambda) = \int d\bfalpha \; \pi(\bfalpha) \; \Theta\left( {L}(\bfalpha) > \lambda \right)
\end{align}
runs over the likelihood values $\lambda$.
Equivalently, the data evidence can be written as
\begin{align}
 p(\bfG_\text{d}|M,\bfsigma,\BK) &= \int_0^1 dX \; \mathcal{L}(X) \notag\\
 &\approx \sum_{n=0}^\infty \Delta X_n  \; \mathcal{L}(X_n) \notag\\
 &\approx  \sum_{n=0}^{n_\text{max}} \Delta X_n \; \myhat{\lambda}_n  \;,
\label{eq:nesa}
\end{align}
where the integral is approximated by the Riemann sum and $\Delta X_n = X_n - X_{n+1}$.
The likelihood $\mathcal{L}(X)$ is a monotonically decreasing function of the prior mass, and the computation is as complicated as the original evaluation of the data evidence.
Skilling proposed a stochastic approach to sample $\mathcal{L}(X)$ based on order statistics
providing the likelihood minima $\{ \myhat{\lambda}_n\}$.
The pseudo code is given in Algorithm \ref{algo_nest}.
\begin{widetext}
\begin{center}
\begin{pseudocode}[ruled]{NESA algorithm}{\{\myhat{\lambda}_{n}\},\{\myhat{\bfalpha}_n\},n_{\text{max}}} \label{algo_nest}
  \textbf{input parameters: } N_\text{w}, \epsilon_{\lambda} \\
 \textbf{initialize } \myhat{\lambda}_{0}=0,\: n=0, \: \\
 \text{draw $N_\text{w}$ configurations}\: \{\bfalpha_{i}\}\: \text{ at random from } \pi(\bfalpha\CS \myhat{\lambda}_{0}) \text{~(Eq. \ref{eq:prior_res})}\\
 \text{take the smallest likelihood value $\myhat{\lambda}=\min\left\{ \lambda_{i}={L}(\bfalpha_i)\right\}$ and its configuration $\myhat{\bfalpha}$} \\
 \textbf{set } \myhat{\lambda}_{n=1}:= \myhat{\lambda} \text{ and } \myhat{\bfalpha}_{n=1}:= \myhat{\bfalpha}\\
 \WHILE \left| \left(\myhat{\lambda}_{n+1}-\myhat{\lambda}_{n}\right)/\myhat{\lambda}_{n+1}\right| \: > \: \epsilon_{\lambda} \DO
 \BEGIN 
  n \GETS n+1\\ 
  \text{replace configuration } \myhat{\bfalpha}
 \text{ with a new configuration drawn from }\: \pi(\bfalpha\CS \myhat{\lambda}_{n})\\
 \text{determine the smallest likelihood  $\myhat{\lambda}=\min\left\{ \lambda_{i}={L}(\bfalpha_i)\right\}$ and its configuration $\myhat{\bfalpha}$}\\
\textbf{set } \myhat{\lambda}_{n+1}=\myhat{\lambda} \text{ and } \myhat{\bfalpha}_{n+1}=\myhat{\bfalpha}
 \END\\
 \textbf{set } n = n_{\text{max}} \\
 \RETURN{\{\myhat{\lambda}_{n}\},\{\myhat{\bfalpha}_n\},n_{\text{max}}}
\end{pseudocode}
\end{center}
\end{widetext}
The nested sampling moves in configuration space ensure
that even well-separated peaks of the likelihood function are sampled correctly.
The crucial step for the NESA algorithm is to draw from
the constrained prior probability, 
\begin{align} \label{eq:prior_res}
 \pi(\bfalpha\CS \myhat{\lambda}_{n}) = \frac{\pi(\bfalpha)}{X(\myhat{\lambda}_{n})} \:\Theta (L(\bfalpha)\:>\: \myhat{\lambda}_{n})\;.
\end{align}
This probability density represents the normalized prior restricted to areas, where $L(\bfalpha)$ exceeds the $\myhat{\lambda}_n$ threshold.
In the present paper, $\pi(\bfalpha)$ is constant within the prior constraints according to Eq. \ref{eq:prior}.
Therefore, we need to draw samples from the uniform distribution constrained by both the likelihood and the prior.
A simple way to draw a sample from Eq. \ref{eq:prior_res} is to clone an existing  configuration, which obviously fulfills all constraints, and 
perform an ordinary Markov chain Monte Carlo update obeying $\pi(\mathbf{x}\CS \myhat{\lambda}_{n})$.
We implemented local updates in the parameters and monitored autocorrelations, which can become considerable depending on the problem.

The prior masses can be derived using order statistics as shown in detail in Ref. \cite{vonderLinden_Bayes_2014}.
We can write the $n$th prior mass as $X_n=\prod_{\nu=1}^{n}\theta_\nu$, where the shrinking factors $\theta_\nu$ are independent and identically distributed random variables
and obey the first-order statistic of the uniform probability density function (PDF), the $\beta$ distribution $p(\theta)=\theta^{N_\text{w}-1}/N_\text{w}$.
Knowing the distribution of $X_n$ and, therefore, of $\Delta X_n$, allows to calculate the expectation value and variance of the Riemann sum in Eq. \ref{eq:nesa} given the set of likelihood minima $\{\myhat{\lambda}_n\}$
obtained from Algorithm \ref{algo_nest}.

Also posterior samples can be generated from a single NESA run by reusing the samples $\{\bfalpha^*_n\}$ obtained from Algorithm \ref{algo_nest}.
Eq. \ref{eq:nesa} shows that the $n$th NESA step contributes with weight $\Delta X_n \lambda^*_n$ to the Riemann sum for calculating the data evidence.
Therefore, samples from the posterior PDF can be provided by choosing $n$ with the corresponding $\bfalpha^*_n$ according to its weight $p(n) \propto \Delta X_n \lambda^*_n$, e.g., by inversion sampling. With such posterior samples $\{ \bfalpha_\nu \}$, the expectation value of the spectral function can be obtained as $\langle \bfA \rangle = \frac{1}{N_\nu}\sum_{\nu=1}^{N_\nu} \bfA(\bfalpha_\nu)$.
%

There are different improvements of NESA going beyond the algorithm we presented within this section,
which may increase the performance, e.g.,
updating more configurations at once \cite{Sivia_DataAnalysis_1996,vonderLinden_Bayes_2014},
using a parallel version of NESA \cite{Henderson_parallelNESA_2014},
extending the update method in the prior sampling \cite{Skilling_Galilean_2012},
or using the knowledge of the position of the minima obtained by optimization algorithms \cite{Martiniani_supenhNESA_2014}.

%
%


\section{Application I: BPAC}
\label{sec:Application}

In this section, we apply BPAC to an impurity problem closely related the one studied in Ref. \cite{Rumetshofer_CuPc_2019}.
In Sec. \ref{sec:AIM}, we define the AIM, which we solve, subsequently, using three different methods: FTPS, CTQMC + MEM, and CTQMC + BPAC.
Technical details of these methods are given in Sec. \ref{sec:methods}, 
whereas the comparison of the results is given in Sec. \ref{sec:BPAC}.

\subsection{The impurity problem}
\label{sec:AIM}

The Hamiltonian of the isolated multiorbital Anderson impurity with on-site energies $\epsilon_{i\sigma}$ and 
interaction parameters $U_{ij}$ for electrons of different spin and $V_{ij}$ for electrons of the same spin is
\begin{align}
 \opH_\text{AI} = \sum_{i\sigma} \epsilon_{i\sigma}  \opn_{i\sigma} + \frac{1}{2}\sum_{ij\sigma} U_{ij}\opn_{i\sigma}\opn_{j\bar\sigma} + \frac{1}{2}\sum_{i\neq j,\sigma} V_{ij}\opn_{i\sigma}\opn_{j\sigma} \;.
\label{eq:modelH}
\end{align}
Here, $\opn_{i\sigma} = \opa_{i\sigma}^\dagger \opa^{\phantom{\dagger}}_{i\sigma}$ is the particle number operator for orbital $i\in\{1,2\}$ and spin $\sigma\in\{\downarrow,\uparrow\}$ in the second quantization with creation (annihilation) operators $\opa_{i\sigma}^\dagger$ ($\opa_{i\sigma}^{\phantom{\dagger}}$).
{In the  AIM, the impurity is coupled to a bath of non-interacting fermions,}
\begin{align}
    \opH_\text{AIM} = \opH_\text{AI} + \sum_{ik\sigma} \tilde V_{ik }\left( \opa_{i\sigma}^\dagger \opc_{ik\sigma}^{\phantom{\dagger}} + h.c. \right) + \sum_{ik\sigma} \epsilon_{ik} \opn_{ik\sigma}\;.
    \label{eq:H_AIM}
\end{align}
$\opc_{ik\sigma}^\dagger$ ($\opc_{ik\sigma}^{\phantom{\dagger}}$) are the creation (annihilation) operators of the $k$th bath state of orbital $i$ with spin $\sigma$. 
For the on-site energy of the impurity, we use $\epsilon_{i\downarrow}=\epsilon-J$ and $\epsilon_{i\uparrow}=\epsilon$ with $\epsilon = -0.25~$~eV and $J = 50~$meV, and
\begin{align}
 U = \begin{pmatrix}
      \tilde U & \tilde U \\
      \tilde U & \tilde U
     \end{pmatrix} \text{~~~and~~~}
 V = \begin{pmatrix}
      0 & \tilde U \\
      \tilde U & 0
      \end{pmatrix}
\end{align}
with $\tilde U = 0.5~$eV.
The bath parameters $\tilde V_{ik }$ and $\epsilon_{ik}$ are obtained from a flat bath hybridization function,  
\begin{align}\label{eq:bathParameters}
  \Delta_i (\omega)  &\stackrel{!}{=}  \sum_k \frac{ \tilde V_{ik}^2}{\omega+\im 0^+ - \epsilon_{ik}}\;
\end{align}
defined by $-2\Im\left(\Delta_i(\omega)\right) = \Gamma\;\Theta(-1 < \omega < 1)$ with $\Gamma=50~$meV. This set of parameters exhibits a generalized Kondo scenario with symmetry between SU(2) \cite{Haldane_Kondo_1978} and SU(4) \cite{Filippone_SU4Kondo_2014} with the corresponding Kondo temperatures of $T_{\text{SU}(2)}=0.36~$K and $T_{\text{SU}(4)}=20~$K. 
Due to the difference in the on-site energies of the impurity orbitals $J$, this AIM exhibits side peaks close to the Abrikosov-Suhl resonance.
We present the spin-down and spin-up spectral functions of this AIM in Figs. \ref{fig:BAC_dn} and \ref{fig:BAC_up}. 

\subsection{Technical details of the methods}
\label{sec:methods}

We solve the AIM in the imaginary-time domain using the CTQMC solver in hybridization expansion as implemented in the TRIQS library \cite{Werner_CTQMC_2006,triqs_2015,Seth_CTHYB_2016}.
We performed 15 CTQMC runs at $\beta = 400~\text{(eV)}^{-1}$ ($T=29~$K), 
each on 20 node points 
and with {$10^{6}$} measurements.
The difference in the impurity on-site energies $J$ lifts the spin degeneracy but keeps the orbital degeneracy intact and,
%
therefore, the 15 CTQMC runs give 30 independent samples. Based on this sample, we estimate reliable variances $\sigma_{i}$ for the QMC data without having to bother about possible autocorrelations.

In the following, we do not distinguish orbitals anymore and just discuss the spectral functions depending on the spin.
The AC in the present paper is performed with  MEM and BPAC.
In both cases, we use $N_\text{d}=401$ data points on an equally spaced $\tau$ grid for $\tau\in [0, \beta]$ and the same amount of $\omega$ points equally spaced on the interval $\omega\in[-1, 1]$.
We applied the MEM of Ref. \cite{Jarrell_MaxEnt_1996} with an alternative evidence approximation \cite{Linden_MaxEnt_1999} and the preblur formalism \cite{Skilling_MaxEnt_1991}.
BPAC is applied as explained in Sec. \ref{sec:ParameterEstimation} using a $N_\text{w} = 1000$ walker and $\epsilon_\lambda=10^{-5}$.

Additionally, we compare the results with those obtained by the real-time FTPS impurity solver, which does not need any AC, since it calculates the Green's function already on the real axis.
In contrast to the CTQMC solver, the FTPS solver is a zero-temperature method, which has to be considered when comparing the results. 

\subsection{Comparison of the results}
\label{sec:BPAC}

\begin{figure}[h]
\begin{center}
  \includegraphics[width=1.0\columnwidth,angle=0]{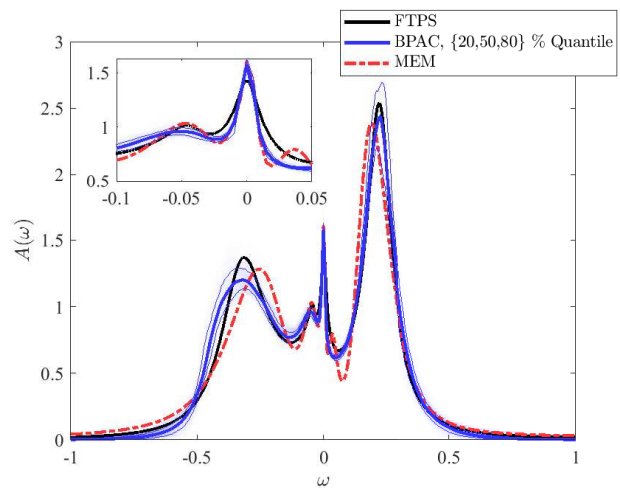}
  \caption{Spectral function for the down electrons obtained by the FTPS solver (black line), CTQMC + BPAC (blue line including confidence intervals), and CTQMC + MEM (red dash-dotted line).
  The BPAC solution does not show the peak slightly above $\omega=0$ of the MEM solution, which is in agreement with the FTPS solution (the inset).}
  \label{fig:BAC_dn}
\end{center}
\end{figure}

First, we employ the FTPS solver for the spin-down part of the AIM and show the corresponding spectrum in Fig. \ref{fig:BAC_dn} (black line).
As expected from the definition of the impurity model in Sec. \ref{sec:AIM}, the spectral function shows Hubbard satellites at approximately $-0.3~$eV and $0.2~$eV and the Abrikosov-Suhl resonance at the chemical potential ($0~$eV).
Additionally, there is a peak at approximately $-50~$meV contributed by the exchange coupling parameter $J$.
The CTQMC + MEM result (red dash-dotted line) shows a spurious peak at approximately $0.04~$eV, which does not appear in the FTPS solution.
To find out whether this peak is supported by the CTQMC data or whether it is an artifact of the AC by MEM, we employ BPAC (blue line).
First, we use a four-peak model $M_4 = M(\bfC = (1, 1, 1, 1) )$ where each peak  consists of a single asymmetric Lorentzian, i.e., with peak complexity $C_n=1$.
The NESA logarithmic data evidence yields $\ln(p(\bfG_\text{d}|M_4,\bfsigma,\BK))=2828.2\pm 0.3$.
Tab. \ref{tab:down} in the App. \ref{sec:App1} shows the prior ranges used and the parameters estimated.
The four-peak BPAC solution (Fig. \ref{fig:BAC_dn}, blue line) does not show the additional peak slightly above $\omega=0$ by construction. 
The evaluation of the five-peak model $M_5 = M(\bfC = (1, 1, 1, 1, 1))$ where we introduce an additional peak at $\mu\in(0.03, 0.07)$
produces a logarithmic data evidence $\ln(p(\bfG_\text{d}|M_5,\bfsigma,\BK))=2825.0\pm 0.3$.
This yields a logarithmic Bayes factor of $3.2\pm 0.6$, 
which corresponds to a probability of $(93\% - 98\%)$ that the four-peak model is preferred over the five-peak model.
This is in agreement with the FTPS solution and demonstrates that the fifth peak at $\sim 0.04~$eV is not supported by the CTQMC data and, therefore, an artifact of the MEM solution.
In general, the spectral function obtained by BPAC depends on the choice of the model $M$.
In the spirit of Bayesian probability theory, we can average over different models weighted by their corresponding model probability.
Therefore, we actually should compute
\begin{align*}
p(\bfA|\bfG_\text{d},\bfsigma,\BK) &= \sum_{i} p(\bfA|\bfG_\text{d},\bfsigma,M_i,\BK) P(M_i|\bfG_\text{d},\bfsigma,\BK) \;.
\end{align*}
If one model is highly preferable, as $M_4$ in the present case, then 
$
p(\bfA|\bfG_\text{d},\bfsigma,\BK) \approx p(\bfA|\bfG_\text{d},\bfsigma,M_4,\BK)
$.
Hence, we plotted $p(\bfA|\bfG_\text{d},\bfsigma,M_4,\BK)$ in Fig. \ref{fig:BAC_dn} (blue line).

\begin{figure}[h]
\begin{center}
  \includegraphics[width=1.0\columnwidth,angle=0]{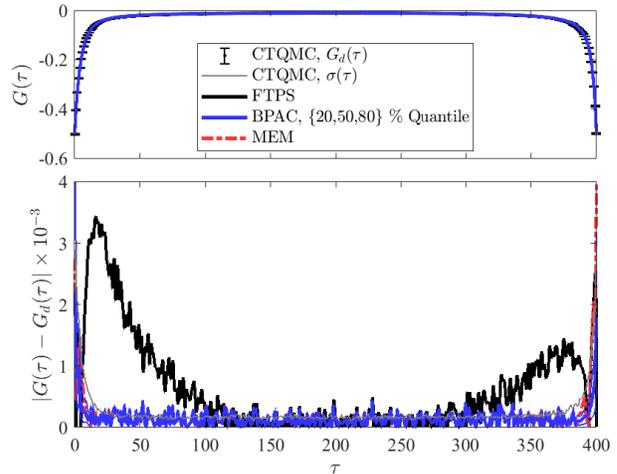}
  \caption{Imaginary-time Green's function (upper panel) and its deviation from the CTQMC data (lower panel) for the spin-down electrons.
  FTPS spectral function ($T=0~$K) (black line)
  shows systematic deviations from the CTQMC data ($T = 29~$K) due to the difference
  in temperature.}
  \label{fig:BAC_dn_G}
\end{center}
\end{figure}

Starting point for the determination of the spectral function $A(\omega)$ are CTQMC data 
on the imaginary-time Green's function $G(\tau)$, which we denote by $G_\text{d}(\tau)$. Inserting the reconstructed spectral function $A(\omega)$  into Eq. \ref{eq:A:omega} yields the reconstructed $G(\tau)$, which allows to asses the misfit in data space. Likewise, we can apply  Eq. \ref{eq:A:omega} to the FTPS spectral function to obtain the corresponding $G(\tau)$.
The reconstructed Green's function for imaginary times $G(\tau)$ is compared with the CTQMC data $G_\text{d}(\tau)$ in Fig. \ref{fig:BAC_dn_G} for the MEM,  BPAC, and FTPS.
Even though the spectral functions of MEM and BPAC differ slightly, $G(\tau)$ of both solutions lies within the error of the CTQMC data.
In the lower panel, the difference between $G(\tau)$ and $G_\text{d}(\tau)$ is shown on an enlarged scale, which reveals a systematic deviation between FTPS and CTQMC data.  
The reason is that the FTPS solver calculates the spectral function at $T=0~$K, whereas  $\beta = 400~(\text{eV})^{-1}$ ($T=29~$K) is used in the CTQMC simulation.

\begin{figure}[h]
\begin{center}
  \includegraphics[width=1.0\columnwidth,angle=0]{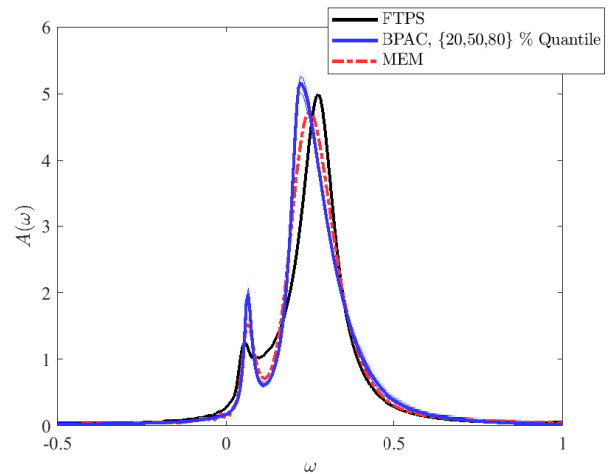}
   \caption{Spectral function for the up electrons obtained by the FTPS solver (black line), CTQMC + BPAC (blue line including confidence intervals), and CTQMC + MEM (red dash-dotted line).
     FTPS calculates the spectral function at $T=0~$K, therefore, there are systematic deviations to the CTQMC + MEM and CTQMC + BPAC solutions obtained at $T=29~$K.}
  \label{fig:BAC_up}
\end{center}
\end{figure}

The spectrum of the spin-up part of the AIM obtained with the FTPS solver is presented in Fig. \ref{fig:BAC_up} (black line) and shows a two-peak structure as does the MEM (red dash-dotted line).
Hence, for parametrizing the spin-up spectral function in BPAC (blue line), we use a two-peak model $M_2 = M(\bfC = (1, 1) )$ with the prior ranges given in Tab. \ref{tab:up} in App. \ref{sec:App1}.
Since we are not interested in specific questions about spurious peaks, we are satisfied with model $M_2$.
%

\section{Application II: MEM + BPAC} 
\label{sec:application2}

In this section, we propose a combination of MEM and BPAC (MEM + BPAC) and apply the method to the impurity problem studied in Ref. \cite{Bauernfeind_FTPS_2017} for SrVO$_{3}$.
We present the details of the AIM in Sec. \ref{sec:AIM2}, give the technical details of MEM + BPAC in Sec. \ref{sec:methods2}, and discuss the results in Sec. \ref{sec:results}.

\subsection{The impurity problem} 
\label{sec:AIM2}

\begin{figure}[h]
\begin{center}
  \includegraphics[width=1.0\columnwidth,angle=0]{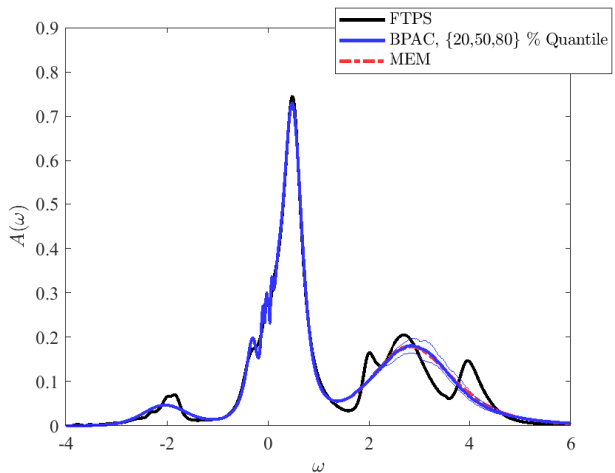}
   \caption{Spectral functions of the AIM for SrVO$_3$ studied in Ref. \cite{Bauernfeind_FTPS_2017}.
  The FTPS solution (black) shows a three-peak structure in the upper Hubbard band, whereas MEM (red dash-dotted) and MEM + BPAC using the three-peak model $M_3$ (blue) do not resolve these peaks. }
  \label{fig:SrVO3}
\end{center}
\end{figure}

The multiorbital AIM discussed in Ref. \cite{Bauernfeind_FTPS_2017} arises from the {\it ab initio} treatment of SrVO$_3$, which has become a test-bed material in DMFT.
The solution of the AIM obtained with the FTPS solver shows a three-peak structure in the upper Hubbard band between $1.75~$eV and $4.25~$eV, see Fig. \ref{fig:SrVO3} (black line).
Ref. \cite{Bauernfeind_FTPS_2017} showed that CTQMC + MEM is not able to resolve these high-energy excitations.
The question we want to address in this section is as follows:
%
%
{
{\it Is the absence of the three-peak structure a failure of MEM, or is it - due to the ill-posed inversion problem - generally impossible to recover certain high-energy details of the spectrum?}}
To answer this question, we applied MEM + BPAC as explained in the following section.

\subsection{Technical details and MEM + BPAC} 
\label{sec:methods2}

In order to obtain an answer to this question, 
we start from the FTPS real-frequency data, transform them to the imaginary-time axis, and add noise on the order of the CTQMC error ($\sigma=10^{-5}$). 
By this procedure, we ensure that we know precisely the error statistics 
of the data, and we know that the correct result has to have the three-peak structure.

We use an inverse temperature $\beta = 200~\text{(eV)}^{-1}$ ($T=58~$K),
 $N_\text{d}=501$ data points on an equally spaced $\tau$ grid in the interval $\tau\in [0, \beta]$ and the same amount of $\omega$ points equally spaced for $\omega\in[-4, 6]$.

Instead of using BPAC as explained in the previous sections, here, we apply a combination of MEM and BPAC. MEM + BPAC takes the MEM solution for a given subinterval of the energy axis and applies BPAC only for the remaining interval.
In that way, the number of parameters is small, which enables faster sampling in the calculation of the evidence integral with NESA.
We take $\Omega\in (1, 1.75)$ as an additional parameter and use the MEM solution for $\omega<\Omega$ and BPAC for $\omega\geq \Omega$. 
We use the prior ranges $1.75<\mu_n< 4.25$ and $0< a_n,\gamma_n^\text{l},\gamma_n^\text{r}< 1$, and $C_n=1$ for each peak $n$.
Furthermore, the remaining parameters describing the right tail are constrained by $4.25<\Omega^r< 6$, $-2<\mu^r< 4.25$, and $1<\nu^r< 10$.
In NESA, we use $N_\text{w} = 2000$ walkers and $\epsilon_\lambda=10^{-5}$.

\subsection{Comparison of the results} 
\label{sec:results}

We applied the MEM of Ref. \cite{Jarrell_MaxEnt_1996} with an alternative evidence approximation \cite{Linden_MaxEnt_1999} and the preblur formalism \cite{Skilling_MaxEnt_1991} and were able to qualitatively reproduce the CTQMC + MEM solution in Fig. 5 of Ref. \cite{Bauernfeind_FTPS_2017}, see Fig. \ref{fig:SrVO3} (red dash-dotted line, mostly covered by the blue line). The MEM spectral function does not show the three-peak structure in the upper Hubbard band.

We applied MEM + BPAC using one-, two-, and three-peak models $M_1$, $M_2$, and $M_3$.
The obtained logarithmic data evidences $\ln(p(\bfG_\text{d}|M_i,\bfsigma,\BK))$ for $i\in\{ 1,2,3 \}$
are $\{5036.7\pm 0.2, 5036.0\pm 0.2, 5035.3\pm 0.2\}$ and correspond to probabilities of 57 \% for $M_1$, 29 \% for $M_2$, and 14 \% for $M_3$. It is interesting to note that the correct three-peak model actually has the lowest probability. 
%
{Still, let us take a look at the result of $M_3$ shown in Fig. \ref{fig:SrVO3}. Surprisingly, the three-peak model looks very similar to the MEM result, i.e., it is not even able to resolve the three-peak structure. Instead, it just shows one large peak in the energy region of the upper Hubbard band.}

\begin{figure}[h]
\begin{center}
  \includegraphics[width=1.0\columnwidth,angle=0]{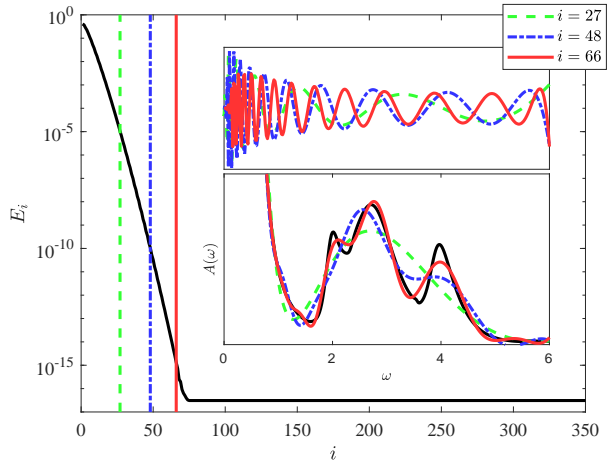}
   \caption{Singular values $E_i$ versus index $i$ for the kernel of the AIM for SrVO$_3$.
   For the selected singular mode indices 27 (green dashed, $E_{27}\approx 10^{-5}$), 48 (blue dash-dotted, $E_{48}\approx 10^{-10}$), and 66 (red, $E_{66}\approx 10^{-15}$), we show the singular modes $\vec v_{i}$ (upper inset) and the projected spectral functions $A(\omega)$ (see Eq. \ref{eq:aux3}) for $N=27,48,66$ (green dashed, blue dash-dotted, and red) (the lower inset).
   Although $66$ singular modes resolve the three-peak structure, $27$ and $48$ do not.}
  \label{fig:SrVO3Kernel}
\end{center}
\end{figure}
%
To elucidate this behavior, we consider the singular value representation of the kernel \cite{Pavarini:852559},
\begin{align}
K &= \sum_{i}E_{i} \; \vec u_{i}  \big( \vec v_{i}\big)^T \;.
\end{align}
Given the vector $\vec A $ of the discretized spectral
function and the corresponding 
vector $\vec G$ of the 
Green's function for discrete imaginary times
as defined in Sec. \ref{sec:AC}, we get
\begin{align}\label{eq:aux0}
\vec G &= K \vec A = \sum_{i} E_{i} \; \vec u_{i} 
\big(\vec v_{i}\cdot \vec A\big)\;.
\end{align}
The misfit defined in Eq. \ref{eq:likelihood} can then be expressed in a very suggestive way.
For simplicity, we assume constant noise $\sigma_{l}=\sigma$, $\forall l$. Then, the misfit to the data vector
$\vec G_\text{d}$ is
\begin{align}\label{eq:aux1}
\chi^{2} 
&= \frac{1}{\sigma^{2}} \sum_{i} \bigg( \big(\vec G_\text{d}\cdot \vec u_{i}\big) - E_{i} 
\big(\vec v_{i}\cdot \vec A\big)\bigg)^{2}\;.
\end{align}
In Fig. \ref{fig:SrVO3Kernel}, the singular values $E_i$ of the kernel matrix are plotted with decreasing magnitude on a logarithmic scale. 
We find that the singular values decrease exponentially and that, above  $i=70$, the singular values are smaller than machine precision.
We select the three singular modes $i=\{27, 48, 66\}$ corresponding to singular values of approximately $\{10^{-5}, 10^{-10}, 10^{-15}\}$, respectively. The corresponding modes $\vec v_{i}$, which are depicted in the upper inset, show an increasing number of nodes with increasing index $i$.
%
%
%
%
{
Minimization of Eq.~\ref{eq:aux1} with respect to the spectral function $\vec{A}$ yields the maximum likelihood solution,
\begin{align}
    \bfA_\text{ML} = \sum_i \frac{\big(\vec G_\text{d}\cdot \vec u_{i}\big)}{E_i} \; \vec v_{i} \;.
\end{align}
Since the singular values decay exponentially, their inverse increases exponentially and small noise in the coefficients 
$\big(\vec G_\text{d}\cdot \vec u_{i}\big)$ becomes amplified.
%
Fig.~\ref{fig:SrVO3Gu} demonstrates nicely that $\big(\vec G_\text{d}\cdot \vec u_{i}\big)$ becomes dominated by noise
above the mode index $i$ where the singular value $E_i$ reaches the order of the noise $\sigma$.
\begin{figure}[h]
\begin{center}
  \includegraphics[width=1.0\columnwidth,angle=0]{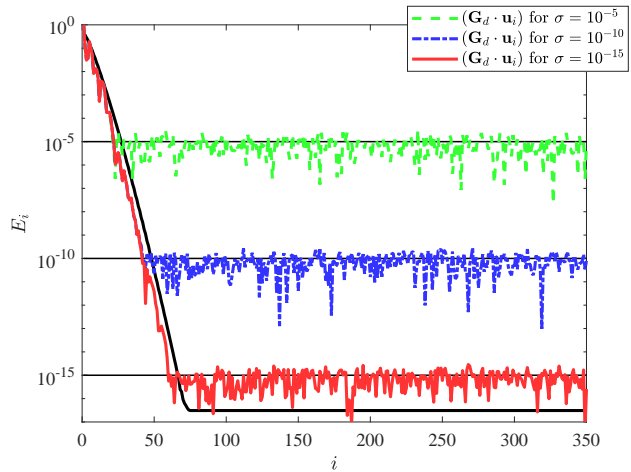}
   \caption{Singular values $E_i$ versus  index $i$ for the kernel of the AIM for SrVO$_3$.
   $\big(\vec G_\text{d}\cdot \vec u_{i}\big)$ is shown for selected noise levels.}
  \label{fig:SrVO3Gu}
\end{center}
\end{figure}}
%
%
%
In order to see which part of the spectrum can, therefore, be reconstructed, we expand $\vec A$ into the mode vectors $\vec v_{i}$, 
\begin{align}\label{eq:aux3}
\vec A &= \sum_{i=1}^{N}     \big(\vec A\cdot \vec v_{i}\big)\;
\vec v_{i} \;.
\end{align}
%
In the lower inset of Fig. \ref{fig:SrVO3Kernel}, we present the projected  FTPS spectrum for $N=27, 48$, and $66$. 
We observe that the three-peak structure is not resolvable at all with $N=27$ and the resulting spectrum (green dashed line in the lower inset) looks similar to the MEM and MEM + BPAC solutions in Fig. \ref{fig:SrVO3}.
$N=48$ allows to resolve two of the three peaks (blue dash-dotted line in the lower inset),
whereas only  $N=66$  resolves the full three-peak structure (red line in the lower inset).
This demonstrates that, for a CTQMC error of $\sigma\geq10^{-10}$, the AC kernel definitely does not allow to resolve the three-peak-structure. CTQMC errors of this magnitude, however, would imply enormous data-acquiring times, and even then only two of the three peaks would be visible.


%
\section{Conclusions} 

We proposed a Bayesian parametric approach for the analytic continuation of noisy imaginary-time Green's function data as, e.g. obtained by CTQMC.
The commonly used Bayesian form-free reconstruction of QMC data is the MEM which is based on the entropic prior that uses a minimum amount of prior information, merely positivity and additivity. 
Due to the nature of the  form-free reconstruction, there are typically as many unknown parameters as noisy data points. This, in combination with the ill-conditioned kernel, can lead to spurious features in the reconstructed 
spectrum. In many applications, however, we have additional prior knowledge, e.g., we know that there will be a small number of peaklike structures of a specific shape. This prior knowledge can be encoded by representing the spectrum in terms of suitably parametrized basis functions, encoding the spectrum with only a few parameters, much less than the number of data points. 
Our approach, which we denote BPAC, employs Bayesian parameter estimation to obtain the parametrized spectral function. 
In the present paper, we used asymmetric Lorentzians and additional tails. 
Of course, in other applications a different basis might be favorable. 

Moreover, we employed Bayesian model comparison to validate different numbers of Lorentzian peaks.
For Bayesian model comparison, the evaluation of high-dimensional evidence integrals is necessary. To this end, we employ NESA, which is particularly efficient for such integration problems.

We demonstrated the capability of BPAC in terms of the CTQMC data for an AIM that has a generalized Kondo scenario and compared the BPAC spectra to the MEM result as well as the spectral function obtained with the real-time FTPS impurity solver.
It was shown that BPAC is able to tell true peaks from artifacts which are present in the MEM's solution close to Abrikosov-Suhl resonance.

In a second application, we studied the AIM arising from the {\it ab initio} treatment of SrVO$_3$. The spectral function obtained with the real-time FTPS solver shows a three-peak structure in the upper Hubbard band, which is not present in the MEM reconstruction of the CTQMC data. Although the MEM cannot resolve the three-peak structure, the rest of the spectrum is captured well. To start with a data set that definitely contains the three-peak structure, we generated imaginary-time data from the real-frequency FTPS spectrum. Adding noise to simulate the CTQMC error, we studied the MEM and BPAC reconstructions of this data set.
To keep the number of parameters and, therefore, the numerical effort small, we employed BPAC focused on the structure in the upper Hubbard band. Therefore, we only described the upper part of the spectrum by Lorentzians whereas keeping the MEM reconstruction for the rest of the spectrum.
Bayesian model comparison then allows to infer  which details of the upper Hubbard band can reliably be inferred from the data. 
{
Considering the singular value decomposition of the kernel allows us to find rigorous arguments how numerical noise is propagated by the kernel. 
} 
Remarkably, we found that the information of the three-peak structure present in the real-frequency spectrum is attenuated by ten orders of magnitude during the transformation to imaginary-time space.
It is, therefore, buried in the noise and impossible to be retrieved from the QMC data.
This means that independent of the model chosen, we obtained a single large peak resembling the MEM solution.
Although BPAC was not able to reconstruct the true shape of the upper Hubbard band, its advantage is that it reliably detects how many details of the spectrum are actually above the noise threshold in the data.

Therefore, we conclude that BPAC is a valuable addition to nonparametric reconstruction methods, such as MEM. The reconstruction can be performed either only with BPAC or with a MEM 
reconstruction that can be used first, and BPAC is employed to assess whether the data support specific features found in the MEM spectral function.


\section{Acknowledgments}

The authors like to thank G. Kraberger and R. Triebl for discussions concerning the maximum entropy method.
{This work was partially supported by the Austrian Science Fund (FWF) through {Grant No. SFB-ViCoM F41P04 and through the START Program No. Y746} as well as by NAWI Graz.}

\section{Appendix}

\subsection{Prior ranges and estimated parameters}
\label{sec:App1}

Tab. \ref{tab:down} shows the prior ranges and the estimated parameters for the four-peak model $M_4$ of the spin-down spectral function of the AIM discussed in Sec. \ref{sec:Application}.
\begin{widetext}
\begin{center}
\begin{table}[h]
  \centering
 \begin{tabular}{l | c | c | c | c}
 & $\mu_n$ & $a_n$ & $\gamma_n^\text{l}$ & $\gamma_n^\text{r}$ \\ \hline\hline
 prior range & (-0.4, -0.2) & (0, 1) & (0.001, 0.3) & (0.001, 0.3) \\
 estimator & -0.32 $\pm$  0.04 & 0.7 $\pm$   0.1 & 0.20 $\pm$  0.06 & 0.18 $\pm$  0.06 \\ \hline
  prior range & (-0.07, -0.03) & (0, 0.5) & (0.001, 0.1) & (0.001, 0.1) \\
 estimator & -0.05 $\pm$  0.01 & 0.08 $\pm$  0.04 & 0.06 $\pm$  0.03 & 0.04 $\pm$  0.02 \\ \hline
   prior range & (-0.02, 0.02) & (0, 0.5) & (0.001, 0.1) & (0.001, 0.1)\\
 estimator & 0.001 $\pm$ 0.003 & 0.022 $\pm$ 0.005 & 0.008 $\pm$ 0.004 & 0.006 $\pm$ 0.003 \\ \hline
   prior range & (0.1, 0.25) & (0, 1) & (0.001, 0.3) & (0.001, 0.3) \\
 estimator & 0.22 $\pm$  0.02 & 0.6 $\pm$   0.1 & 0.06 $\pm$  0.01 & 0.09 $\pm$  0.05 \\ \hline
 \end{tabular} \\
  \begin{tabular}{ l l | c | c | c }
 & & $\limOm$ & $\mu$ & $\nu$ \\ \hline\hline
 left tail & prior range & (-0.5, -0.4) & (-0.4, 0.5) & (1, 10) \\
   & estimator & -0.45 $\pm$  0.03 & 0.0 $\pm$   0.2 & 7 $\pm$     2 \\ \hline
 right tail & prior range & (0.25, 0.5) & (-0.5, 0.25) & (1, 10) \\
   & estimator & 0.36 $\pm$  0.07  & -0.1 $\pm$   0.2 & 7 $\pm$     2 \\ \hline
 \end{tabular} 
  \caption{Prior ranges and estimated parameters for model $M_4$ for the spin-down spectral function.}
  \label{tab:down}
\end{table}  
\end{center}
\end{widetext}

Tab. \ref{tab:up} shows the prior ranges and the estimated parameters for the two-peak model $M_2$ of the spin-up spectral function of the AIM discussed in Sec. \ref{sec:Application}.
\begin{widetext}
\begin{center}
\begin{table}[h]
  \centering
 \begin{tabular}{l | c | c | c | c}
 & $\mu_n$ & $a_n$ & $\gamma_n^\text{l}$ & $\gamma_n^\text{r}$ \\ \hline\hline
 prior range &( 0.03, 0.07)  & (    0,  0.1) & (0.001,  0.1) & (0.001,  0.1) \\
 estimator & 0.063 $\pm$  0.001 & 0.078 $\pm$  0.002 & 0.0116 $\pm$ 0.0004 & 0.016 $\pm$  0.002 \\ \hline
  prior range & (  0.2,  0.3) & (    0,  1.5) & (0.001,  0.5) & (0.001,  0.5) \\
 estimator & 0.218 $\pm$  0.002 & 1.06$\pm$0.03 & 0.0320 $\pm$ 0.0007 &  0.100 $\pm$  0.007 \\ \hline
 \end{tabular} \\
  \begin{tabular}{ l l | c | c | c }
 & & $\limOm$ & $\mu$ & $\nu$ \\ \hline\hline
 left tail & prior range & (    0, 0.03) & ( 0.03,  0.5) & (    1,   10) \\
   & estimator & 0.003 $\pm$  0.001 & 0.1076 $\pm$ 0.0007 & 1.281 $\pm$  0.005 \\ \hline
 right tail & prior range & (  0.3,  0.5) & ( -0.5,  0.3) & (    1,   10) \\
   & estimator & 0.37 $\pm$   0.04 & 0.0 $\pm$    0.2 &  4 $\pm$      2 \\ \hline
 \end{tabular} 
  \caption{Prior ranges and estimated parameters for model $M_2$ for the spin-up spectral function.}
  \label{tab:up}
\end{table}
\end{center}
\end{widetext}

\section*{References}

\bibliographystyle{apsrev4-1}
\bibliography{literature}

\begin{thebibliography}{45}%
\makeatletter
\providecommand \@ifxundefined [1]{%
 \@ifx{#1\undefined}
}%
\providecommand \@ifnum [1]{%
 \ifnum #1\expandafter \@firstoftwo
 \else \expandafter \@secondoftwo
 \fi
}%
\providecommand \@ifx [1]{%
 \ifx #1\expandafter \@firstoftwo
 \else \expandafter \@secondoftwo
 \fi
}%
\providecommand \natexlab [1]{#1}%
\providecommand \enquote  [1]{``#1''}%
\providecommand \bibnamefont  [1]{#1}%
\providecommand \bibfnamefont [1]{#1}%
\providecommand \citenamefont [1]{#1}%
\providecommand \href@noop [0]{\@secondoftwo}%
\providecommand \href [0]{\begingroup \@sanitize@url \@href}%
\providecommand \@href[1]{\@@startlink{#1}\@@href}%
\providecommand \@@href[1]{\endgroup#1\@@endlink}%
\providecommand \@sanitize@url [0]{\catcode `\\12\catcode `\$12\catcode
  `\&12\catcode `\#12\catcode `\^12\catcode `\_12\catcode `\%12\relax}%
\providecommand \@@startlink[1]{}%
\providecommand \@@endlink[0]{}%
\providecommand \url  [0]{\begingroup\@sanitize@url \@url }%
\providecommand \@url [1]{\endgroup\@href {#1}{\urlprefix }}%
\providecommand \urlprefix  [0]{URL }%
\providecommand \Eprint [0]{\href }%
\providecommand \doibase [0]{http://dx.doi.org/}%
\providecommand \selectlanguage [0]{\@gobble}%
\providecommand \bibinfo  [0]{\@secondoftwo}%
\providecommand \bibfield  [0]{\@secondoftwo}%
\providecommand \translation [1]{[#1]}%
\providecommand \BibitemOpen [0]{}%
\providecommand \bibitemStop [0]{}%
\providecommand \bibitemNoStop [0]{.\EOS\space}%
\providecommand \EOS [0]{\spacefactor3000\relax}%
\providecommand \BibitemShut  [1]{\csname bibitem#1\endcsname}%
\let\auto@bib@innerbib\@empty
\bibitem [{\citenamefont {Hirsch}\ and\ \citenamefont
  {Fye}(1986)}]{Hirsch_QMC_1986}%
  \BibitemOpen
  \bibfield  {author} {\bibinfo {author} {\bibfnamefont {J.~E.}\ \bibnamefont
  {Hirsch}}\ and\ \bibinfo {author} {\bibfnamefont {R.~M.}\ \bibnamefont
  {Fye}},\ }\href {\doibase 10.1103/PhysRevLett.56.2521} {\bibfield  {journal}
  {\bibinfo  {journal} {Phys. Rev. Lett.}\ }\textbf {\bibinfo {volume} {56}},\
  \bibinfo {pages} {2521} (\bibinfo {year} {1986})}\BibitemShut {NoStop}%
\bibitem [{\citenamefont {Foulkes}\ \emph {et~al.}(2001)\citenamefont
  {Foulkes}, \citenamefont {Mitas}, \citenamefont {Needs},\ and\ \citenamefont
  {Rajagopal}}]{Foulkes_QMC_2001}%
  \BibitemOpen
  \bibfield  {author} {\bibinfo {author} {\bibfnamefont {W.~M.~C.}\
  \bibnamefont {Foulkes}}, \bibinfo {author} {\bibfnamefont {L.}~\bibnamefont
  {Mitas}}, \bibinfo {author} {\bibfnamefont {R.~J.}\ \bibnamefont {Needs}}, \
  and\ \bibinfo {author} {\bibfnamefont {G.}~\bibnamefont {Rajagopal}},\ }\href
  {\doibase 10.1103/RevModPhys.73.33} {\bibfield  {journal} {\bibinfo
  {journal} {Rev. Mod. Phys.}\ }\textbf {\bibinfo {volume} {73}},\ \bibinfo
  {pages} {33} (\bibinfo {year} {2001})}\BibitemShut {NoStop}%
\bibitem [{\citenamefont {Werner}\ \emph {et~al.}(2006)\citenamefont {Werner},
  \citenamefont {Comanac}, \citenamefont {de' Medici}, \citenamefont {Troyer},\
  and\ \citenamefont {Millis}}]{Werner_CTQMC_2006}%
  \BibitemOpen
  \bibfield  {author} {\bibinfo {author} {\bibfnamefont {P.}~\bibnamefont
  {Werner}}, \bibinfo {author} {\bibfnamefont {A.}~\bibnamefont {Comanac}},
  \bibinfo {author} {\bibfnamefont {L.}~\bibnamefont {de' Medici}}, \bibinfo
  {author} {\bibfnamefont {M.}~\bibnamefont {Troyer}}, \ and\ \bibinfo {author}
  {\bibfnamefont {A.~J.}\ \bibnamefont {Millis}},\ }\href {\doibase
  10.1103/PhysRevLett.97.076405} {\bibfield  {journal} {\bibinfo  {journal}
  {Phys. Rev. Lett.}\ }\textbf {\bibinfo {volume} {97}},\ \bibinfo {pages}
  {076405} (\bibinfo {year} {2006})}\BibitemShut {NoStop}%
\bibitem [{\citenamefont {Gull}\ \emph {et~al.}(2011)\citenamefont {Gull},
  \citenamefont {Millis}, \citenamefont {Lichtenstein}, \citenamefont
  {Rubtsov}, \citenamefont {Troyer},\ and\ \citenamefont
  {Werner}}]{Gull_CTQMC_2011}%
  \BibitemOpen
  \bibfield  {author} {\bibinfo {author} {\bibfnamefont {E.}~\bibnamefont
  {Gull}}, \bibinfo {author} {\bibfnamefont {A.~J.}\ \bibnamefont {Millis}},
  \bibinfo {author} {\bibfnamefont {A.~I.}\ \bibnamefont {Lichtenstein}},
  \bibinfo {author} {\bibfnamefont {A.~N.}\ \bibnamefont {Rubtsov}}, \bibinfo
  {author} {\bibfnamefont {M.}~\bibnamefont {Troyer}}, \ and\ \bibinfo {author}
  {\bibfnamefont {P.}~\bibnamefont {Werner}},\ }\href {\doibase
  10.1103/RevModPhys.83.349} {\bibfield  {journal} {\bibinfo  {journal} {Rev.
  Mod. Phys.}\ }\textbf {\bibinfo {volume} {83}},\ \bibinfo {pages} {349}
  (\bibinfo {year} {2011})}\BibitemShut {NoStop}%
\bibitem [{\citenamefont {Ferris-Prabhu}\ and\ \citenamefont
  {Withers}(1973)}]{FERRISPRABHU_Pade_1973}%
  \BibitemOpen
  \bibfield  {author} {\bibinfo {author} {\bibfnamefont {A.}~\bibnamefont
  {Ferris-Prabhu}}\ and\ \bibinfo {author} {\bibfnamefont {D.}~\bibnamefont
  {Withers}},\ }\href {\doibase https://doi.org/10.1016/0021-9991(73)90127-7}
  {\bibfield  {journal} {\bibinfo  {journal} {Journal of Computational
  Physics}\ }\textbf {\bibinfo {volume} {13}},\ \bibinfo {pages} {94 }
  (\bibinfo {year} {1973})}\BibitemShut {NoStop}%
\bibitem [{\citenamefont {Arsenault}\ \emph {et~al.}(2017)\citenamefont
  {Arsenault}, \citenamefont {Neuberg}, \citenamefont {Hannah},\ and\
  \citenamefont {Millis}}]{Arsenault_machinelearning_2017}%
  \BibitemOpen
  \bibfield  {author} {\bibinfo {author} {\bibfnamefont {L.-F.}\ \bibnamefont
  {Arsenault}}, \bibinfo {author} {\bibfnamefont {R.}~\bibnamefont {Neuberg}},
  \bibinfo {author} {\bibfnamefont {L.~A.}\ \bibnamefont {Hannah}}, \ and\
  \bibinfo {author} {\bibfnamefont {A.~J.}\ \bibnamefont {Millis}},\ }\href
  {\doibase 10.1088/1361-6420/aa8d93} {\bibfield  {journal} {\bibinfo
  {journal} {Inverse Problems}\ }\textbf {\bibinfo {volume} {33}},\ \bibinfo
  {pages} {115007} (\bibinfo {year} {2017})}\BibitemShut {NoStop}%
\bibitem [{\citenamefont {White}\ \emph {et~al.}(1991)\citenamefont {White},
  \citenamefont {Landau}, \citenamefont {Mon},\ and\ \citenamefont
  {Schüttler}}]{White_AverSpectrumMethod_1991}%
  \BibitemOpen
  \bibfield  {author} {\bibinfo {author} {\bibfnamefont {S.}~\bibnamefont
  {White}}, \bibinfo {author} {\bibfnamefont {D.}~\bibnamefont {Landau}},
  \bibinfo {author} {\bibfnamefont {K.}~\bibnamefont {Mon}}, \ and\ \bibinfo
  {author} {\bibfnamefont {B.}~\bibnamefont {Schüttler}},\ }\href@noop {}
  {\emph {\bibinfo {title} {Computer Simulation Studies in Condensed Matter
  Physics III}}}\ (\bibinfo {year} {1991})\ pp.\ \bibinfo {pages}
  {145--153}\BibitemShut {NoStop}%
\bibitem [{\citenamefont {Sandvik}(1998)}]{Sandvik_StochAC_1998}%
  \BibitemOpen
  \bibfield  {author} {\bibinfo {author} {\bibfnamefont {A.~W.}\ \bibnamefont
  {Sandvik}},\ }\href {\doibase 10.1103/PhysRevB.57.10287} {\bibfield
  {journal} {\bibinfo  {journal} {Phys. Rev. B}\ }\textbf {\bibinfo {volume}
  {57}},\ \bibinfo {pages} {10287} (\bibinfo {year} {1998})}\BibitemShut
  {NoStop}%
\bibitem [{\citenamefont {Mishchenko}\ \emph {et~al.}(2000)\citenamefont
  {Mishchenko}, \citenamefont {Prokof'ev}, \citenamefont {Sakamoto},\ and\
  \citenamefont {Svistunov}}]{Mishchenko_samplingMaxEnt_2000}%
  \BibitemOpen
  \bibfield  {author} {\bibinfo {author} {\bibfnamefont {A.~S.}\ \bibnamefont
  {Mishchenko}}, \bibinfo {author} {\bibfnamefont {N.~V.}\ \bibnamefont
  {Prokof'ev}}, \bibinfo {author} {\bibfnamefont {A.}~\bibnamefont {Sakamoto}},
  \ and\ \bibinfo {author} {\bibfnamefont {B.~V.}\ \bibnamefont {Svistunov}},\
  }\href {\doibase 10.1103/PhysRevB.62.6317} {\bibfield  {journal} {\bibinfo
  {journal} {Phys. Rev. B}\ }\textbf {\bibinfo {volume} {62}},\ \bibinfo
  {pages} {6317} (\bibinfo {year} {2000})}\BibitemShut {NoStop}%
\bibitem [{\citenamefont {{Beach}}(2004)}]{Beach_sochasticAC_2004}%
  \BibitemOpen
  \bibfield  {author} {\bibinfo {author} {\bibfnamefont {K.~S.~D.}\
  \bibnamefont {{Beach}}},\ }\href@noop {} {\bibfield  {journal} {\bibinfo
  {journal} {eprint arXiv:cond-mat/0403055}\ } (\bibinfo {year}
  {2004})}\BibitemShut {NoStop}%
\bibitem [{\citenamefont {Fuchs}\ \emph {et~al.}(2010)\citenamefont {Fuchs},
  \citenamefont {Pruschke},\ and\ \citenamefont
  {Jarrell}}]{Fuchs_samplingMaxEnt_2010}%
  \BibitemOpen
  \bibfield  {author} {\bibinfo {author} {\bibfnamefont {S.}~\bibnamefont
  {Fuchs}}, \bibinfo {author} {\bibfnamefont {T.}~\bibnamefont {Pruschke}}, \
  and\ \bibinfo {author} {\bibfnamefont {M.}~\bibnamefont {Jarrell}},\ }\href
  {\doibase 10.1103/PhysRevE.81.056701} {\bibfield  {journal} {\bibinfo
  {journal} {Phys. Rev. E}\ }\textbf {\bibinfo {volume} {81}},\ \bibinfo
  {pages} {056701} (\bibinfo {year} {2010})}\BibitemShut {NoStop}%
\bibitem [{\citenamefont {Sandvik}(2016)}]{Sandvik_StochAC_2016}%
  \BibitemOpen
  \bibfield  {author} {\bibinfo {author} {\bibfnamefont {A.~W.}\ \bibnamefont
  {Sandvik}},\ }\href {\doibase 10.1103/PhysRevE.94.063308} {\bibfield
  {journal} {\bibinfo  {journal} {Phys. Rev. E}\ }\textbf {\bibinfo {volume}
  {94}},\ \bibinfo {pages} {063308} (\bibinfo {year} {2016})}\BibitemShut
  {NoStop}%
\bibitem [{\citenamefont {Ghanem}(2017)}]{Ghanem_StochS_2017}%
  \BibitemOpen
  \bibfield  {author} {\bibinfo {author} {\bibfnamefont {K.}~\bibnamefont
  {Ghanem}},\ }\emph {\bibinfo {title} {{S}tochastic {A}nalytic {C}ontinuation:
  {A} {B}ayesian {A}pproach}},\ \href {\doibase 10.18154/RWTH-2017-06704}
  {\bibinfo {type} {Dissertation}},\ \bibinfo  {school} {RWTH Aachen
  University}, \bibinfo {address} {Aachen} (\bibinfo {year} {2017}),\ \bibinfo
  {note} {veröffentlicht auf dem Publikationsserver der RWTH Aachen
  University; Dissertation, RWTH Aachen University, 2017}\BibitemShut {NoStop}%
\bibitem [{\citenamefont {Silver}\ \emph {et~al.}(1990)\citenamefont {Silver},
  \citenamefont {Sivia},\ and\ \citenamefont
  {Gubernatis}}]{Silver_CTQMCMaxEnt_1990}%
  \BibitemOpen
  \bibfield  {author} {\bibinfo {author} {\bibfnamefont {R.~N.}\ \bibnamefont
  {Silver}}, \bibinfo {author} {\bibfnamefont {D.~S.}\ \bibnamefont {Sivia}}, \
  and\ \bibinfo {author} {\bibfnamefont {J.~E.}\ \bibnamefont {Gubernatis}},\
  }\href {\doibase 10.1103/PhysRevB.41.2380} {\bibfield  {journal} {\bibinfo
  {journal} {Phys. Rev. B}\ }\textbf {\bibinfo {volume} {41}},\ \bibinfo
  {pages} {2380} (\bibinfo {year} {1990})}\BibitemShut {NoStop}%
\bibitem [{\citenamefont {Gubernatis}\ \emph {et~al.}(1991)\citenamefont
  {Gubernatis}, \citenamefont {Jarrell}, \citenamefont {Silver},\ and\
  \citenamefont {Sivia}}]{Gubernatis_QMCMaxEnt_1991}%
  \BibitemOpen
  \bibfield  {author} {\bibinfo {author} {\bibfnamefont {J.~E.}\ \bibnamefont
  {Gubernatis}}, \bibinfo {author} {\bibfnamefont {M.}~\bibnamefont {Jarrell}},
  \bibinfo {author} {\bibfnamefont {R.~N.}\ \bibnamefont {Silver}}, \ and\
  \bibinfo {author} {\bibfnamefont {D.~S.}\ \bibnamefont {Sivia}},\ }\href
  {\doibase 10.1103/PhysRevB.44.6011} {\bibfield  {journal} {\bibinfo
  {journal} {Phys. Rev. B}\ }\textbf {\bibinfo {volume} {44}},\ \bibinfo
  {pages} {6011} (\bibinfo {year} {1991})}\BibitemShut {NoStop}%
\bibitem [{\citenamefont {Gull}\ and\ \citenamefont
  {Skilling}(1984)}]{Skilling_MaxEnt_1984}%
  \BibitemOpen
  \bibfield  {author} {\bibinfo {author} {\bibfnamefont {S.}~\bibnamefont
  {Gull}}\ and\ \bibinfo {author} {\bibfnamefont {J.}~\bibnamefont
  {Skilling}},\ }\href@noop {} {\bibfield  {journal} {\bibinfo  {journal} {IEE
  Proceedings F (Communications, Radar and Signal Processing)}\ }\textbf
  {\bibinfo {volume} {131}},\ \bibinfo {pages} {646} (\bibinfo {year}
  {1984})}\BibitemShut {NoStop}%
\bibitem [{\citenamefont {Bao}\ \emph {et~al.}(2016)\citenamefont {Bao},
  \citenamefont {Tang}, \citenamefont {Summers}, \citenamefont {Zhang},
  \citenamefont {Webster}, \citenamefont {Scarola},\ and\ \citenamefont
  {Maier}}]{Bao_optimMaxEnt_2016}%
  \BibitemOpen
  \bibfield  {author} {\bibinfo {author} {\bibfnamefont {F.}~\bibnamefont
  {Bao}}, \bibinfo {author} {\bibfnamefont {Y.}~\bibnamefont {Tang}}, \bibinfo
  {author} {\bibfnamefont {M.}~\bibnamefont {Summers}}, \bibinfo {author}
  {\bibfnamefont {G.}~\bibnamefont {Zhang}}, \bibinfo {author} {\bibfnamefont
  {C.}~\bibnamefont {Webster}}, \bibinfo {author} {\bibfnamefont
  {V.}~\bibnamefont {Scarola}}, \ and\ \bibinfo {author} {\bibfnamefont
  {T.~A.}\ \bibnamefont {Maier}},\ }\href {\doibase 10.1103/PhysRevB.94.125149}
  {\bibfield  {journal} {\bibinfo  {journal} {Phys. Rev. B}\ }\textbf {\bibinfo
  {volume} {94}},\ \bibinfo {pages} {125149} (\bibinfo {year}
  {2016})}\BibitemShut {NoStop}%
\bibitem [{\citenamefont {Kraberger}\ \emph {et~al.}(2017)\citenamefont
  {Kraberger}, \citenamefont {Triebl}, \citenamefont {Zingl},\ and\
  \citenamefont {Aichhorn}}]{Kraberger_MaxEnt_2017}%
  \BibitemOpen
  \bibfield  {author} {\bibinfo {author} {\bibfnamefont {G.~J.}\ \bibnamefont
  {Kraberger}}, \bibinfo {author} {\bibfnamefont {R.}~\bibnamefont {Triebl}},
  \bibinfo {author} {\bibfnamefont {M.}~\bibnamefont {Zingl}}, \ and\ \bibinfo
  {author} {\bibfnamefont {M.}~\bibnamefont {Aichhorn}},\ }\href {\doibase
  10.1103/PhysRevB.96.155128} {\bibfield  {journal} {\bibinfo  {journal} {Phys.
  Rev. B}\ }\textbf {\bibinfo {volume} {96}},\ \bibinfo {pages} {155128}
  (\bibinfo {year} {2017})}\BibitemShut {NoStop}%
\bibitem [{\citenamefont {von der Linden. Volker Dose. Udo~von
  Toussaint}(2014)}]{vonderLinden_Bayes_2014}%
  \BibitemOpen
  \bibfield  {author} {\bibinfo {author} {\bibfnamefont {W.}~\bibnamefont {von
  der Linden. Volker Dose. Udo~von Toussaint}},\ }\href@noop {} {\emph
  {\bibinfo {title} {Bayesian Probability Theory}}}\ (\bibinfo  {publisher}
  {Cambridge University Press},\ \bibinfo {year} {2014})\BibitemShut {NoStop}%
\bibitem [{\citenamefont {Gregory}(2005)}]{Gregory_Bayes_2005}%
  \BibitemOpen
  \bibfield  {author} {\bibinfo {author} {\bibfnamefont {P.}~\bibnamefont
  {Gregory}},\ }\href@noop {} {\emph {\bibinfo {title} {Bayesian Logical Data
  Analysis for the Physical Sciences: A Comparative Approach with
  Mathematica{\textregistered} Support}}}\ (\bibinfo  {publisher} {Cambridge
  University Press},\ \bibinfo {year} {2005})\BibitemShut {NoStop}%
\bibitem [{\citenamefont {Jaynes}\ \emph {et~al.}(2003)\citenamefont {Jaynes},
  \citenamefont {Jaynes}, \citenamefont {Bretthorst},\ and\ \citenamefont
  {Press}}]{Jaynes_LogicOfSience_2003}%
  \BibitemOpen
  \bibfield  {author} {\bibinfo {author} {\bibfnamefont {E.}~\bibnamefont
  {Jaynes}}, \bibinfo {author} {\bibfnamefont {E.}~\bibnamefont {Jaynes}},
  \bibinfo {author} {\bibfnamefont {G.}~\bibnamefont {Bretthorst}}, \ and\
  \bibinfo {author} {\bibfnamefont {C.~U.}\ \bibnamefont {Press}},\ }\href@noop
  {} {\emph {\bibinfo {title} {Probability Theory: The Logic of Science}}}\
  (\bibinfo  {publisher} {Cambridge University Press},\ \bibinfo {year}
  {2003})\BibitemShut {NoStop}%
\bibitem [{\citenamefont {Sivia}\ and\ \citenamefont
  {Skilling}(2006)}]{Sivia_DataAnalysis_1996}%
  \BibitemOpen
  \bibfield  {author} {\bibinfo {author} {\bibfnamefont {D.}~\bibnamefont
  {Sivia}}\ and\ \bibinfo {author} {\bibfnamefont {J.}~\bibnamefont
  {Skilling}},\ }\href@noop {} {\emph {\bibinfo {title} {Data Analysis: A
  Bayesian Tutorial}}},\ Oxford science publications\ (\bibinfo  {publisher}
  {OUP Oxford},\ \bibinfo {year} {2006})\BibitemShut {NoStop}%
\bibitem [{\citenamefont {Skilling}(2004)}]{Skilling_NESA_2004}%
  \BibitemOpen
  \bibfield  {author} {\bibinfo {author} {\bibfnamefont {J.}~\bibnamefont
  {Skilling}},\ }\href {\doibase 10.1063/1.1835238} {\bibfield  {journal}
  {\bibinfo  {journal} {AIP Conference Proceedings}\ }\textbf {\bibinfo
  {volume} {735}},\ \bibinfo {pages} {395} (\bibinfo {year}
  {2004})}\BibitemShut {NoStop}%
\bibitem [{\citenamefont {Parcollet}\ \emph {et~al.}(2015)\citenamefont
  {Parcollet}, \citenamefont {Ferrero}, \citenamefont {Ayral}, \citenamefont
  {Hafermann}, \citenamefont {Krivenko}, \citenamefont {Messio},\ and\
  \citenamefont {Seth}}]{triqs_2015}%
  \BibitemOpen
  \bibfield  {author} {\bibinfo {author} {\bibfnamefont {O.}~\bibnamefont
  {Parcollet}}, \bibinfo {author} {\bibfnamefont {M.}~\bibnamefont {Ferrero}},
  \bibinfo {author} {\bibfnamefont {T.}~\bibnamefont {Ayral}}, \bibinfo
  {author} {\bibfnamefont {H.}~\bibnamefont {Hafermann}}, \bibinfo {author}
  {\bibfnamefont {I.}~\bibnamefont {Krivenko}}, \bibinfo {author}
  {\bibfnamefont {L.}~\bibnamefont {Messio}}, \ and\ \bibinfo {author}
  {\bibfnamefont {P.}~\bibnamefont {Seth}},\ }\href {\doibase
  https://doi.org/10.1016/j.cpc.2015.04.023} {\bibfield  {journal} {\bibinfo
  {journal} {Computer Physics Communications}\ }\textbf {\bibinfo {volume}
  {196}},\ \bibinfo {pages} {398 } (\bibinfo {year} {2015})}\BibitemShut
  {NoStop}%
\bibitem [{\citenamefont {Seth}\ \emph {et~al.}(2016)\citenamefont {Seth},
  \citenamefont {Krivenko}, \citenamefont {Ferrero},\ and\ \citenamefont
  {Parcollet}}]{Seth_CTHYB_2016}%
  \BibitemOpen
  \bibfield  {author} {\bibinfo {author} {\bibfnamefont {P.}~\bibnamefont
  {Seth}}, \bibinfo {author} {\bibfnamefont {I.}~\bibnamefont {Krivenko}},
  \bibinfo {author} {\bibfnamefont {M.}~\bibnamefont {Ferrero}}, \ and\
  \bibinfo {author} {\bibfnamefont {O.}~\bibnamefont {Parcollet}},\ }\href
  {\doibase https://doi.org/10.1016/j.cpc.2015.10.023} {\bibfield  {journal}
  {\bibinfo  {journal} {Computer Physics Communications}\ }\textbf {\bibinfo
  {volume} {200}},\ \bibinfo {pages} {274 } (\bibinfo {year}
  {2016})}\BibitemShut {NoStop}%
\bibitem [{\citenamefont {Bauernfeind}\ \emph {et~al.}(2017)\citenamefont
  {Bauernfeind}, \citenamefont {Zingl}, \citenamefont {Triebl}, \citenamefont
  {Aichhorn},\ and\ \citenamefont {Evertz}}]{Bauernfeind_FTPS_2017}%
  \BibitemOpen
  \bibfield  {author} {\bibinfo {author} {\bibfnamefont {D.}~\bibnamefont
  {Bauernfeind}}, \bibinfo {author} {\bibfnamefont {M.}~\bibnamefont {Zingl}},
  \bibinfo {author} {\bibfnamefont {R.}~\bibnamefont {Triebl}}, \bibinfo
  {author} {\bibfnamefont {M.}~\bibnamefont {Aichhorn}}, \ and\ \bibinfo
  {author} {\bibfnamefont {H.~G.}\ \bibnamefont {Evertz}},\ }\href {\doibase
  10.1103/PhysRevX.7.031013} {\bibfield  {journal} {\bibinfo  {journal} {Phys.
  Rev. X}\ }\textbf {\bibinfo {volume} {7}},\ \bibinfo {pages} {031013}
  (\bibinfo {year} {2017})}\BibitemShut {NoStop}%
\bibitem [{\citenamefont {Bauernfeind}(2018)}]{Bauernfeind_THESIS_2018}%
  \BibitemOpen
  \bibfield  {author} {\bibinfo {author} {\bibfnamefont {D.}~\bibnamefont
  {Bauernfeind}},\ }\emph {\bibinfo {title} {Fork Tensor Product States}},\
  \href@noop {} {\bibinfo {type} {Phd thesis}},\ \bibinfo  {school} {Graz
  University of Technology} (\bibinfo {year} {2018})\BibitemShut {NoStop}%
\bibitem [{\citenamefont {Metzner}\ and\ \citenamefont
  {Vollhardt}(1989)}]{MetznerVollhardt_Dinf}%
  \BibitemOpen
  \bibfield  {author} {\bibinfo {author} {\bibfnamefont {W.}~\bibnamefont
  {Metzner}}\ and\ \bibinfo {author} {\bibfnamefont {D.}~\bibnamefont
  {Vollhardt}},\ }\href {\doibase 10.1103/PhysRevLett.62.324} {\bibfield
  {journal} {\bibinfo  {journal} {Phys. Rev. Lett.}\ }\textbf {\bibinfo
  {volume} {62}},\ \bibinfo {pages} {324} (\bibinfo {year} {1989})}\BibitemShut
  {NoStop}%
\bibitem [{\citenamefont {Georges}\ and\ \citenamefont
  {Kotliar}(1992)}]{Georges_DMFToriginal}%
  \BibitemOpen
  \bibfield  {author} {\bibinfo {author} {\bibfnamefont {A.}~\bibnamefont
  {Georges}}\ and\ \bibinfo {author} {\bibfnamefont {G.}~\bibnamefont
  {Kotliar}},\ }\href {\doibase 10.1103/PhysRevB.45.6479} {\bibfield  {journal}
  {\bibinfo  {journal} {Phys. Rev. B}\ }\textbf {\bibinfo {volume} {45}},\
  \bibinfo {pages} {6479} (\bibinfo {year} {1992})}\BibitemShut {NoStop}%
\bibitem [{\citenamefont {Skilling}(1989)}]{Skilling_MaxEnt_1989}%
  \BibitemOpen
  \bibfield  {author} {\bibinfo {author} {\bibfnamefont {J.}~\bibnamefont
  {Skilling}},\ }\enquote {\bibinfo {title} {Classic maximum entropy},}\ in\
  \href {\doibase 10.1007/978-94-015-7860-8_3} {\emph {\bibinfo {booktitle}
  {Maximum Entropy and Bayesian Methods}}},\ \bibinfo {editor} {edited by\
  \bibinfo {editor} {\bibfnamefont {J.}~\bibnamefont {Skilling}}}\ (\bibinfo
  {publisher} {Springer Netherlands},\ \bibinfo {address} {Dordrecht},\
  \bibinfo {year} {1989})\ pp.\ \bibinfo {pages} {45--52}\BibitemShut {NoStop}%
\bibitem [{\citenamefont {{Gull S.F.}}\ and\ \citenamefont {{Daniell
  G.J.}}(1978)}]{Gull_MaxEnt_1978}%
  \BibitemOpen
  \bibfield  {author} {\bibinfo {author} {\bibnamefont {{Gull S.F.}}}\ and\
  \bibinfo {author} {\bibnamefont {{Daniell G.J.}}},\ }\href {\doibase
  https://doi.org/10.1038/272686a0} {\bibfield  {journal} {\bibinfo  {journal}
  {Nature}\ }\textbf {\bibinfo {volume} {272}},\ \bibinfo {pages} {686–690}
  (\bibinfo {year} {1978})}\BibitemShut {NoStop}%
\bibitem [{\citenamefont {Gull}(1989)}]{Gull_MaxEnt_1989}%
  \BibitemOpen
  \bibfield  {author} {\bibinfo {author} {\bibfnamefont {S.~F.}\ \bibnamefont
  {Gull}},\ }\enquote {\bibinfo {title} {Developments in maximum entropy data
  analysis},}\ in\ \href {\doibase 10.1007/978-94-015-7860-8_4} {\emph
  {\bibinfo {booktitle} {Maximum Entropy and Bayesian Methods}}},\ \bibinfo
  {editor} {edited by\ \bibinfo {editor} {\bibfnamefont {J.}~\bibnamefont
  {Skilling}}}\ (\bibinfo  {publisher} {Springer Netherlands},\ \bibinfo
  {address} {Dordrecht},\ \bibinfo {year} {1989})\ pp.\ \bibinfo {pages}
  {53--71}\BibitemShut {NoStop}%
\bibitem [{\citenamefont {Fischer}\ \emph {et~al.}(1996)\citenamefont
  {Fischer}, \citenamefont {Linden},\ and\ \citenamefont
  {Dose}}]{fischer_importance_1996}%
  \BibitemOpen
  \bibfield  {author} {\bibinfo {author} {\bibfnamefont {R.}~\bibnamefont
  {Fischer}}, \bibinfo {author} {\bibfnamefont {W.~V.~D.}\ \bibnamefont
  {Linden}}, \ and\ \bibinfo {author} {\bibfnamefont {V.}~\bibnamefont
  {Dose}},\ }in\ \href {\doibase 10.1007/978-94-011-5430-7_27} {\emph {\bibinfo
  {booktitle} {Maximum {Entropy} and {Bayesian} {Methods}}}},\ \bibinfo {series
  and number} {\bibinfo {series} {Fundamental {Theories} of {Physics}}\
  No.~\bibinfo {number} {79}},\ \bibinfo {editor} {edited by\ \bibinfo {editor}
  {\bibfnamefont {K.~M.}\ \bibnamefont {Hanson}}\ and\ \bibinfo {editor}
  {\bibfnamefont {R.~N.}\ \bibnamefont {Silver}}}\ (\bibinfo  {publisher}
  {Springer Netherlands},\ \bibinfo {year} {1996})\ pp.\ \bibinfo {pages}
  {229--236}\BibitemShut {NoStop}%
\bibitem [{\citenamefont {Bryan}(1990)}]{Bryan_MaxEnt_1990}%
  \BibitemOpen
  \bibfield  {author} {\bibinfo {author} {\bibfnamefont {R.~K.}\ \bibnamefont
  {Bryan}},\ }\enquote {\bibinfo {title} {Solving oversampled data problems by
  maximum entropy},}\ in\ \href {\doibase 10.1007/978-94-009-0683-9_12} {\emph
  {\bibinfo {booktitle} {Maximum Entropy and Bayesian Methods}}},\ \bibinfo
  {editor} {edited by\ \bibinfo {editor} {\bibfnamefont {P.~F.}\ \bibnamefont
  {Foug{\`e}re}}}\ (\bibinfo  {publisher} {Springer Netherlands},\ \bibinfo
  {address} {Dordrecht},\ \bibinfo {year} {1990})\ pp.\ \bibinfo {pages}
  {221--232}\BibitemShut {NoStop}%
\bibitem [{\citenamefont {Skilling}(2006)}]{Skilling_NESA_2006}%
  \BibitemOpen
  \bibfield  {author} {\bibinfo {author} {\bibfnamefont {J.}~\bibnamefont
  {Skilling}},\ }\href {\doibase 10.1214/06-BA127} {\bibfield  {journal}
  {\bibinfo  {journal} {Bayesian Anal.}\ }\textbf {\bibinfo {volume} {1}},\
  \bibinfo {pages} {833} (\bibinfo {year} {2006})}\BibitemShut {NoStop}%
\bibitem [{\citenamefont {Henderson}\ and\ \citenamefont
  {Goggans}(2014)}]{Henderson_parallelNESA_2014}%
  \BibitemOpen
  \bibfield  {author} {\bibinfo {author} {\bibfnamefont {R.~W.}\ \bibnamefont
  {Henderson}}\ and\ \bibinfo {author} {\bibfnamefont {P.~M.}\ \bibnamefont
  {Goggans}},\ }\href {\doibase 10.1063/1.4903717} {\bibfield  {journal}
  {\bibinfo  {journal} {AIP Conference Proceedings}\ }\textbf {\bibinfo
  {volume} {1636}},\ \bibinfo {pages} {100} (\bibinfo {year}
  {2014})}\BibitemShut {NoStop}%
\bibitem [{\citenamefont {Skilling}(2012)}]{Skilling_Galilean_2012}%
  \BibitemOpen
  \bibfield  {author} {\bibinfo {author} {\bibfnamefont {J.}~\bibnamefont
  {Skilling}},\ }\href {\doibase 10.1063/1.3703630} {\bibfield  {journal}
  {\bibinfo  {journal} {AIP Conference Proceedings}\ }\textbf {\bibinfo
  {volume} {1443}},\ \bibinfo {pages} {145} (\bibinfo {year}
  {2012})}\BibitemShut {NoStop}%
\bibitem [{\citenamefont {Martiniani}\ \emph {et~al.}(2014)\citenamefont
  {Martiniani}, \citenamefont {Stevenson}, \citenamefont {Wales},\ and\
  \citenamefont {Frenkel}}]{Martiniani_supenhNESA_2014}%
  \BibitemOpen
  \bibfield  {author} {\bibinfo {author} {\bibfnamefont {S.}~\bibnamefont
  {Martiniani}}, \bibinfo {author} {\bibfnamefont {J.~D.}\ \bibnamefont
  {Stevenson}}, \bibinfo {author} {\bibfnamefont {D.~J.}\ \bibnamefont
  {Wales}}, \ and\ \bibinfo {author} {\bibfnamefont {D.}~\bibnamefont
  {Frenkel}},\ }\href {\doibase 10.1103/PhysRevX.4.031034} {\bibfield
  {journal} {\bibinfo  {journal} {Phys. Rev. X}\ }\textbf {\bibinfo {volume}
  {4}},\ \bibinfo {pages} {031034} (\bibinfo {year} {2014})}\BibitemShut
  {NoStop}%
\bibitem [{\citenamefont {Rumetshofer}\ \emph {et~al.}(2019)\citenamefont
  {Rumetshofer}, \citenamefont {Bauernfeind}, \citenamefont {Arrigoni},\ and\
  \citenamefont {von~der Linden}}]{Rumetshofer_CuPc_2019}%
  \BibitemOpen
  \bibfield  {author} {\bibinfo {author} {\bibfnamefont {M.}~\bibnamefont
  {Rumetshofer}}, \bibinfo {author} {\bibfnamefont {D.}~\bibnamefont
  {Bauernfeind}}, \bibinfo {author} {\bibfnamefont {E.}~\bibnamefont
  {Arrigoni}}, \ and\ \bibinfo {author} {\bibfnamefont {W.}~\bibnamefont
  {von~der Linden}},\ }\href {\doibase 10.1103/PhysRevB.99.045148} {\bibfield
  {journal} {\bibinfo  {journal} {Phys. Rev. B}\ }\textbf {\bibinfo {volume}
  {99}},\ \bibinfo {pages} {045148} (\bibinfo {year} {2019})}\BibitemShut
  {NoStop}%
\bibitem [{\citenamefont {Haldane}(1978)}]{Haldane_Kondo_1978}%
  \BibitemOpen
  \bibfield  {author} {\bibinfo {author} {\bibfnamefont {F.~D.~M.}\
  \bibnamefont {Haldane}},\ }\href {\doibase 10.1103/PhysRevLett.40.416}
  {\bibfield  {journal} {\bibinfo  {journal} {Phys. Rev. Lett.}\ }\textbf
  {\bibinfo {volume} {40}},\ \bibinfo {pages} {416} (\bibinfo {year}
  {1978})}\BibitemShut {NoStop}%
\bibitem [{\citenamefont {Filippone}\ \emph {et~al.}(2014)\citenamefont
  {Filippone}, \citenamefont {Moca}, \citenamefont {Zar\'and},\ and\
  \citenamefont {Mora}}]{Filippone_SU4Kondo_2014}%
  \BibitemOpen
  \bibfield  {author} {\bibinfo {author} {\bibfnamefont {M.}~\bibnamefont
  {Filippone}}, \bibinfo {author} {\bibfnamefont {C.~P.}\ \bibnamefont {Moca}},
  \bibinfo {author} {\bibfnamefont {G.}~\bibnamefont {Zar\'and}}, \ and\
  \bibinfo {author} {\bibfnamefont {C.}~\bibnamefont {Mora}},\ }\href {\doibase
  10.1103/PhysRevB.90.121406} {\bibfield  {journal} {\bibinfo  {journal} {Phys.
  Rev. B}\ }\textbf {\bibinfo {volume} {90}},\ \bibinfo {pages} {121406(R)}
  (\bibinfo {year} {2014})}\BibitemShut {NoStop}%
\bibitem [{\citenamefont {Jarrell}\ and\ \citenamefont
  {Gubernatis}(1996)}]{Jarrell_MaxEnt_1996}%
  \BibitemOpen
  \bibfield  {author} {\bibinfo {author} {\bibfnamefont {M.}~\bibnamefont
  {Jarrell}}\ and\ \bibinfo {author} {\bibfnamefont {J.}~\bibnamefont
  {Gubernatis}},\ }\href {\doibase
  https://doi.org/10.1016/0370-1573(95)00074-7} {\bibfield  {journal} {\bibinfo
   {journal} {Physics Reports}\ }\textbf {\bibinfo {volume} {269}},\ \bibinfo
  {pages} {133 } (\bibinfo {year} {1996})}\BibitemShut {NoStop}%
\bibitem [{\citenamefont {von~der Linden}\ \emph {et~al.}(1999)\citenamefont
  {von~der Linden}, \citenamefont {Preuss},\ and\ \citenamefont
  {Dose}}]{Linden_MaxEnt_1999}%
  \BibitemOpen
  \bibfield  {author} {\bibinfo {author} {\bibfnamefont {W.}~\bibnamefont
  {von~der Linden}}, \bibinfo {author} {\bibfnamefont {R.}~\bibnamefont
  {Preuss}}, \ and\ \bibinfo {author} {\bibfnamefont {V.}~\bibnamefont
  {Dose}},\ }in\ \href@noop {} {\emph {\bibinfo {booktitle} {Maximum Entropy
  and Bayesian Methods}}},\ \bibinfo {editor} {edited by\ \bibinfo {editor}
  {\bibfnamefont {W.}~\bibnamefont {von~der Linden}}, \bibinfo {editor}
  {\bibfnamefont {V.}~\bibnamefont {Dose}}, \bibinfo {editor} {\bibfnamefont
  {R.}~\bibnamefont {Fischer}}, \ and\ \bibinfo {editor} {\bibfnamefont
  {R.}~\bibnamefont {Preuss}}}\ (\bibinfo  {publisher} {Springer Netherlands},\
  \bibinfo {address} {Dordrecht},\ \bibinfo {year} {1999})\ pp.\ \bibinfo
  {pages} {319--326}\BibitemShut {NoStop}%
\bibitem [{\citenamefont {Skilling}(1991)}]{Skilling_MaxEnt_1991}%
  \BibitemOpen
  \bibfield  {author} {\bibinfo {author} {\bibfnamefont {J.}~\bibnamefont
  {Skilling}},\ }\enquote {\bibinfo {title} {Fundamentals of maxent in data
  analysis},}\ in\ \href@noop {} {\emph {\bibinfo {booktitle} {Maximum Entropy
  in Action}}},\ \bibinfo {editor} {edited by\ \bibinfo {editor} {\bibfnamefont
  {B.}~\bibnamefont {Buck}}\ and\ \bibinfo {editor} {\bibfnamefont
  {V.}~\bibnamefont {Macaulay}}}\ (\bibinfo  {publisher} {Clarendon Press},\
  \bibinfo {address} {Oxford},\ \bibinfo {year} {1991})\ pp.\ \bibinfo {pages}
  {19--40}\BibitemShut {NoStop}%
\bibitem [{Pav(2018)}]{Pavarini:852559}%
  \BibitemOpen
  \href {http://juser.fz-juelich.de/record/852559} {\emph {\bibinfo {title}
  {{DMFT}: {F}rom {I}nfinite {D}imensions to {R}eal {M}aterials}}},\ \bibinfo
  {series} {Modeling and Simulation}, Vol.~\bibinfo {volume} {8},\ \bibinfo
  {organization} {Autumn School on Correlated Electrons, Juelich (Germany), 17
  Sep 2018 - 21 Sep 2018}\ (\bibinfo  {publisher} {Forschungszentrum Juelich
  GmbH Zentralbibliothek, Velag},\ \bibinfo {address} {Juelich},\ \bibinfo
  {year} {2018})\BibitemShut {NoStop}%
\end{thebibliography}%

\end{document}